\documentclass[twocolumn]{aastex63}

\usepackage{xcolor}

\usepackage{amsmath}

\newcommand{\CfA}{\affiliation{Center for Astrophysics \textbar{} Harvard \& Smithsonian, 60 Garden Street, Cambridge, MA 02138-1516, USA}}
\newcommand{\IAIFI}{\affiliation{The NSF AI Institute for Artificial Intelligence and Fundamental Interactions, USA}}
\newcommand{\LCO}{\affiliation{Las Cumbres Observatory, 6740 Cortona Drive, Suite 102, Goleta, CA 93117-5575, USA}}
\newcommand{\UCSB}{\affiliation{Department of Physics, University of California, Santa Barbara, CA 93106-9530, USA}}

\newcommand{\STSci}{\affiliation{Space Telescope Science Institute, 3700 San Martin Dr, Baltimore, MD 21218, USA}}

\newcommand{\TAU}{\affiliation{School of Physics and Astronomy, Tel Aviv University, Tel Aviv 69978, Israel}}
\newcommand{\CIFAR}{\affiliation{CIFAR Azrieli Global Scholars program, CIFAR, 661 University Ave., Toronto, ON, M5G 1M1, Canada}}
\newcommand{\NAOJ}{\affiliation{National Astronomical Observatory of Japan, National Institutes of Natural Sciences, 2-21-1 Osawa, Mitaka, Tokyo 181-8588, Japan}}

\newcommand{\IPMU}{\affiliation{Kavli Institute for the Physics and Mathematics of the Universe (WPI), The University of Tokyo Institutes for Advanced Study, The University of Tokyo, 5-1-5 Kashiwanoha, Kashiwa, Chiba 277-8583, Japan}}

\newcommand{\Columbia}{\affiliation{Department of Physics and Columbia Astrophysics Laboratory, Columbia University, 538 West 120th Street, New York, NY 10027, USA }}
\newcommand{\CCA}{\affiliation{Center for Computational Astrophysics, Flatiron Institute, New York, NY 10010, USA}}
\newcommand{\SOKEN}{\affiliation{Department of Astronomical Science, School of Physical Sciences, The Graduate University of Advanced Studies (SOKENDAI), 2-21-1 Osawa, Mitaka, Tokyo 181-8588, Japan}}
\newcommand{\Konan}{\affiliation{Department of Physics, Faculty of Science and Engineering, Konan University, 8-9-1 Okamoto, Kobe, Hyogo 658-8501, Japan}}

\newcommand{\McGill}{\affiliation{Department of Physics, McGill University, 3600 rue University, Montr\'eal, QC, H3A 2T8, Canada}}
\newcommand{\McGillSpace}{\affiliation{McGill Space Institute, McGill University, 3550 rue University, Montr\'eal, QC, H3A 2A7, Canada}}

\received{2022 November 7}
\revised{2022 December 17}
\accepted{2022 December 22}
\published{2023 April 26}
\submitjournal{ApJL}

\shorttitle{Limits on Optical Emission from Well-localized FRBs}
\shortauthors{Hiramatsu et al.}

\begin{document}

\title{\bf \Large Limits on Simultaneous and Delayed Optical Emission from Well-localized Fast Radio Bursts}

\author[0000-0002-1125-9187]{Daichi Hiramatsu}
\CfA\IAIFI

\correspondingauthor{Daichi~Hiramatsu}
\email{daichi.hiramatsu@cfa.harvard.edu}

\author[0000-0002-9392-9681]{Edo~Berger}
\CfA\IAIFI

\author[0000-0002-4670-7509]{Brian~D.~Metzger}
\Columbia\CCA

\author[0000-0001-6395-6702]{Sebastian~Gomez}
\STSci

\author[0000-0001-6637-5401]{Allyson~Bieryla}
\CfA

\author[0000-0001-7090-4898]{Iair~Arcavi}
\TAU\CIFAR

\author[0000-0003-4253-656X]{D.~Andrew~Howell}
\LCO\UCSB

\author[0000-0001-7348-6900]{Ryan~Mckinven}
\McGill\McGillSpace

\author[0000-0001-8537-3153]{Nozomu~Tominaga}
\NAOJ\SOKEN\Konan\IPMU

\begin{abstract}
We present the largest compilation to date of optical observations during and following fast radio bursts (FRBs). The data set includes our dedicated {\it simultaneous} and follow-up observations, as well as serendipitous archival survey observations, for a sample of 15 well-localized FRBs: eight repeating and seven one-off sources. Our simultaneous (and nearly simultaneous with a $0.4$ s delay) optical observations of 13 (1) bursts from the repeating FRB~20220912A provide the deepest such limits to date for any extragalactic FRB, reaching a luminosity limit of $\nu L_\nu\lesssim 10^{42}$ erg s$^{-1}$ ($\lesssim 2\times10^{41}$ erg s$^{-1}$) with $15$--$400$ s exposures; an optical-flux-to-radio-fluence ratio of $f_{\rm opt}/F_{\rm radio}\lesssim 10^{-7}$ ms$^{-1}$ ($\lesssim 10^{-8}$ ms$^{-1}$); and flux ratio of $f_{\rm opt}/f_{\rm radio}\lesssim 0.02$--$\lesssim 2\times 10^{-5}$ ($\lesssim 10^{-6}$) on millisecond to second timescales.  These simultaneous limits provide useful constraints in the context of FRB emission models, such as the pulsar magnetosphere and pulsar nebula models. Interpreting all available optical limits in the context of the synchrotron maser model, we find that they constrain the flare energies to $\lesssim 10^{43}$--$10^{49}$ erg (depending on the distances of the various repeating FRBs, with $\lesssim 10^{39}$ erg for the Galactic SGR~1935+2154). These limits are generally at least an order of magnitude larger than those inferred from the FRBs themselves, although in the case of FRB~20220912A our simultaneous and rapid follow-up observations severely restrict the model parameter space. We conclude by exploring the potential of future simultaneous and rapid-response observations with large optical telescopes.
\end{abstract}

\keywords{
\href{https://vocabs.ardc.edu.au/repository/api/lda/aas/the-unified-astronomy-thesaurus/current/resource.html?uri=http://astrothesaurus.org/uat/2008}{Radio transient sources (2008)}; 
\href{https://vocabs.ardc.edu.au/repository/api/lda/aas/the-unified-astronomy-thesaurus/current/resource.html?uri=http://astrothesaurus.org/uat/1851}{Transient sources (1851)}; 
\href{https://vocabs.ardc.edu.au/repository/api/lda/aas/the-unified-astronomy-thesaurus/current/resource.html?uri=http://astrothesaurus.org/uat/992}{Magnetars (992)}; 
\href{https://vocabs.ardc.edu.au/repository/api/lda/aas/the-unified-astronomy-thesaurus/current/resource.html?uri=http://astrothesaurus.org/uat/1306}{Pulsars (1306)}
}

\section{Introduction} 
\label{sec:intro}

Fast radio bursts (FRBs) are incredibly bright, millisecond-duration pulses at GHz frequencies (e.g., \citealt{Lorimer2007Sci...318..777L}; see \citealt{Petroff2019A&ARv..27....4P,Petroff2022A&ARv..30....2P} for reviews). Their dispersion measures (DMs), the integrated electron column density along the line of sight, exceed the range of the Milky Way (MW) plus its halo, implying an extragalactic origin. While most FRBs appear as one-off events, a growing subset are known to repeat (e.g., \citealt{Spitler2016Natur.531..202S}; see \citealt{CHIME/FRB2019ApJ...885L..24C} and \citealt{Fonseca2020ApJ...891L...6F} for sample studies), and it is possible that the entire population repeats on a wide range of timescales (e.g., \citealt{Nicholl2017ApJ...843...84N,Ravi2019NatAs...3..928R,Cui2021MNRAS.500.3275C}). To date, about 20 FRBs (both one-off and repeaters) have been precisely localized through their radio emission, providing an initial, though potentially biased, view of their host-galaxy environments (see, e.g., \citealt{Chatterjee2017Natur.541...58C, Marcote2017ApJ...834L...8M,Tendulkar2017ApJ...834L...7T}; see further \citealt{Heintz2020ApJ...903..152H,Bhandari2022AJ....163...69B}), and directly confirming their extragalactic origin.


Despite the rapidly increasing sample size and the availability of some localizations and host-galaxy properties, the physical origin(s) and mechanism(s) of FRBs remain unknown, with dozens of proposed sources and emission mechanisms published to date (see, e.g., \citealt{Platts2019PhR...821....1P} for a living list of theoretical FRB models).\footnote{\url{https://frbtheorycat.org/index.php/Main_Page}} Given the short duration and noncatastrophic nature of the repeating FRBs, models involving young neutron stars or black holes are particularly popular; the recent faint FRB-like detection from the Galactic magnetar SGR~1935+2154 may support such a picture \citep{CHIME/FRB2020Natur.587...54C,Bochenek2020Natur.587...59B}. Some of the precisely localized extragalactic FRBs (e.g., FRB~20200120E in an M81 globular cluster; \citealt{Bhardwaj2021ApJ...910L..18B,Kirsten2022Natur.602..585K}) show no correlation with star formation, possibly indicating an origin from older neutron stars, accretion-powered binaries (e.g., \citealt{Sridhar+21}), or from young magnetars born from an older progenitor channel (e.g., accretion-induced collapse of white dwarfs or the merger of two neutron stars; \citealt{Margalit2019ApJ...886..110M, Kremer+22}). 

A significant barrier to our understanding of FRBs is their sole detection in the radio band (see, e.g., \citealt{Chen2020ApJ...897..146C,Nicastro2021Univ....7...76N} for reviews on multiwavelength observations).\footnote{Temporally coincident X-ray bursts from the Galactic SGR~1935+2154 \citep{Mereghetti2020ApJ...898L..29M,Li2021NatAs...5..378L,Ridnaia2021NatAs...5..372R,Tavani2021NatAs...5..401T} remain the only nonradio transient detections. However, such bursts would not be detectable in extragalactic FRBs due to their much larger distances (see e.g., \citealt{Scholz2016ApJ...833..177S,Scholz2017ApJ...846...80S,Scholz2020ApJ...901..165S,Tavani2020ApJ...893L..42T} for X-ray nondetections of extragalactic FRBs).} This situation is reminiscent of the first two decades of gamma-ray burst (GRB) research when only gamma-ray emission had been detected; it was only through rapid multiwavelength detections of associated afterglows that the progenitors and physics of GRBs were eventually uncovered.\footnote{A key distinction is that for GRBs detections of afterglows at longer wavelengths were critical for precise localizations and the identification of host galaxies (and hence a distance scale and stellar population properties), whereas FRBs can be precisely localized directly in the radio band.}

In the same vein, rapid and deep optical follow-up of FRBs may shed light on their physical mechanisms. FRB models predict a wide range of possible luminosities and timescales for optical transient counterparts, ranging from no emission at all, to luminous ($\gtrsim 10^{41}$\,erg\,s$^{-1}$) afterglows on a millisecond timescale, to fainter ($\lesssim 10^{39}$\,erg\,s$^{-1}$) afterglows on timescales of minutes to hours (e.g., \citealt{Lyubarsky2014MNRAS.442L...9L,Beloborodov2017ApJ...843L..26B,Metzger2019MNRAS.485.4091M,Yang2019ApJ...878...89Y,Margalit2020ApJ...899L..27M, Margalit2020MNRAS.494.4627M,Beloborodov2020ApJ...896..142B}). Previous optical follow-up attempts have suffered from the combination of FRB poor localizations, large distances, delayed announcements of bursts, and a limited sample of repeating FRBs, resulting in only weak constraints (e.g., $\lesssim10^{45}$\,erg\,s$^{-1}$ over a millisecond timescale to $\lesssim10^{43}$\,erg\,s$^{-1}$ over a minute timescale; \citealt{Hardy2017MNRAS.472.2800H, MAGIC2018MNRAS.481.2479M, Andreoni2020ApJ...896L...2A, Niino2022ApJ...931..109N}); dedicated monitoring of a single burst from the well-localized repeating FRB~20180916B ($d_L\approx 150$\,Mpc; \citealt{Marcote2020Natur.577..190M}) placed marginal constraints ($\lesssim10^{40}$\,erg\,s$^{-1}$ over minute timescale) on models such as the synchrotron maser with high burst energies and circumburst densities \citep{Kilpatrick2021ApJ...907L...3K}. 

Here, we report the largest set of optical constraints to date, for a sample of eight well-localized repeating FRBs (including the Galactic SGR~1935+2154) and seven well-localized non-repeating FRBs, using an extensive set of archival data, as well as our own dedicated {\it follow-up} observations of the nearby repeating FRB~20200120E, and {\it simultaneous} observations of the newly discovered and highly active FRB~20220912A \citep{McKinven2022ATel15679....1M}, the most sensitive simultaneous observations to date for any extragalactic FRB. The paper is structured as follows. We summarize the FRB sample, optical observations, and data reduction in Section~\ref{sec:data}. In Section~\ref{sec:ana}, we analyze and discuss the optical luminosity, flux, and fluence limits in the context of the FRBs' radio properties, and compare these to theoretical models. We summarize our findings and conclude with a future outlook in Section~\ref{sec:sum}.

\section{Sample and Observations} 
\label{sec:data}

\subsection{Fast Radio Burst Sample} 
\label{sec:sample}

\begin{deluxetable*}{cccccccccccc}
\tabletypesize{\scriptsize}
\tablecaption{The Sample of Well-localized Repeating FRBs \label{tab:FRBsample}}
\tablehead{
\colhead{FRB} & \colhead{R.A.} & \colhead{Decl.} & \colhead{Redshift} & \colhead{$d_L$\tablenotemark{a}} & \colhead{$A_{V,{\rm MW}}$\tablenotemark{b}} & \colhead{Events\tablenotemark{c}} & \colhead{Frequency} & \colhead{DM} & \colhead{Width\tablenotemark{d}} & \colhead{Flux\tablenotemark{d}} & \colhead{Fluence\tablenotemark{d}} \\[-6pt]
\colhead{} & \colhead{(deg)} & \colhead{(deg)} & \colhead{} & \colhead{(Mpc)} & \colhead{(mag)} & \colhead{(\#)} & \colhead{(GHz)} & \colhead{(pc\,cm$^{-3}$)} & \colhead{(ms)} & \colhead{(Jy)} & \colhead{(Jy\,ms)} 
}
\startdata
20200120E & 149.4779140(3)\tablenotemark{e} & +68.8169036(4)\tablenotemark{e} & $-0.00013$\tablenotemark{f} & $3.6$\tablenotemark{f} & $0.200$ & 74\tablenotemark{g} & $0.40$--$2.3$ & $87.7$--$88$ & $0.014$--$0.70$ & $0.10$--$60$ & $0.04$--$2$ \\
20180916B & 29.5031257(6)\tablenotemark{h} & +65.7167542(6)\tablenotemark{h} & $0.0337$\tablenotemark{h} & $149$\tablenotemark{h} & $2.712$ & 244\tablenotemark{i} &  $0.12$--$4.9$ & $343$--$356$ & $0.3$--$158$ & $0.12$--$20$ & $0.08$--$300$ \\
20220912A & 347.2704(6)\tablenotemark{j} & +48.7071(3)\tablenotemark{j} & $0.077$\tablenotemark{j} & $344$ & $0.637$ & 72\tablenotemark{k} & $0.11$--$2.3$ & $218$--$228$ & $0.8$--$300$ & $3$--$290$ & $1.5$--$900$ \\
20201124A & 77.0146142(8)\tablenotemark{l} & +26.0606959(7)\tablenotemark{l} & $0.0979$\tablenotemark{l} & $444$ & $1.964$ & 2914\tablenotemark{m} & $0.40$--$2.3$ & $409$--$424$ & $0.9$--$300$ & $0.004$--$220$ & $0.005$--$770$ \\
20121102A & 82.994575(3)\tablenotemark{n} & +33.147940(1)\tablenotemark{n} & $0.1927$\tablenotemark{n} & $927$ & $2.098$ & 3658\tablenotemark{o} & $0.60$--$7.5$ & $527$--$698$ & $0.01$--$78.52$ & $0.002$--$70$ & $0.002$--$35$\\
20190520B & 240.51780(3)\tablenotemark{p} & $-$11.28814(2)\tablenotemark{p} & $0.241$\tablenotemark{p} & $1192$ & $0.769$ & 230\tablenotemark{q} & $1.4$--$6.2$ & $1164$--$1291$ &  $0.7$--$33.1$ & $0.002$--$2$ & $0.029$--$6$ \\
20180301A & 93.2268(2)\tablenotemark{r} & +4.6711(2)\tablenotemark{r} & $0.3304$\tablenotemark{r} & $1712$ & $1.231$ & 22\tablenotemark{s} & $1.2$--$1.3$ & $510$--$536$ & $1.7$--$12.3$ & $0.005$--$1.2$ & $0.021$--$4.90$ \\
\hline
SGR & & & & (kpc) & \\
\hline
1935+2154 & 293.7317(2)\tablenotemark{t} & +21.8966(2)\tablenotemark{t} & Galactic\tablenotemark{t} & $9.0$\tablenotemark{u} & $7.2$\tablenotemark{v} &  14\tablenotemark{w} & $0.11$--$5.6$ & $313$--$333$ & $0.22$--$2000$ & $0.030$--$2.5\times10^6$ & $0.06$--$1.5\times10^6$  \\
\enddata

\tablecomments{Measurements of each individual burst are published in its entirety in machine-readable form.}

\tablenotetext{a}{Calculated from the host redshift assuming a standard Lambda cold dark matter ($\Lambda$CDM) cosmology with $H_0=71.0$\, km\,s$^{-1}$\,Mpc$^{-1}$, $\Omega_{\Lambda}=0.7$, and $\Omega_m=0.3$, unless otherwise noted.}

\tablenotetext{b}{From \cite{Schlafly2011ApJ...737..103S}, retrieved via the NASA/IPAC Infrared Science Archive (IRSA), unless otherwise noted.}

\tablenotetext{c}{Only the bursts with reported time of arrival are included.}

\tablenotetext{d}{If only two of the width ($t_{\rm FRB}$), flux ($f_{\rm radio}$), and fluence ($F_{\rm radio}$) are reported for a particular burst, the third parameter is estimated assuming $F_{\rm radio} \sim f_{\rm radio}t_{\rm FRB}$.}

\tablenotetext{e}{Best localized with the European VLBI Network (EVN; \citealt{Kirsten2022Natur.602..585K}).}

\tablenotetext{f}{From \cite{Speights2012ApJ...752...52S}, retrieved via the NASA/IPAC Extragalactic Database.}

\tablenotetext{g}{Data sources: CHIME (CHIME/FRB Public Database, \citealt{Bhardwaj2021ApJ...910L..18B}), the Deep Sapce Network (DSN; \citealt{Majid2021ApJ...919L...6M}), EVN \citep{Kirsten2022Natur.602..585K}, Effelsberg \citep{Nimmo2023MNRAS.520.2281N}.}

\tablenotetext{h}{Best localized with EVN, and the host spectroscopic redshift measured with Gemini-North \citep{Marcote2020Natur.577..190M}.}

\tablenotetext{i}{Data sources: CHIME (CHIME/FRB Public Database; \citealt{CHIME/FRB2019ApJ...885L..24C, CHIME/FRB2020Natur.582..351C, Pleunis2021ApJ...911L...3P}), Apertif \citep{Pastor-Marazuela2021Natur.596..505P}, Effelsberg \citep{Bethapudi2022arXiv220713669B}, EVN \citep{Marcote2020Natur.577..190M}, Green Bank Telescope (GBT; \citealt{Chawla2020ApJ...896L..41C, Sand2022ApJ...932...98S}), Low Frequency Array (LOFAR; \citealt{Pastor-Marazuela2021Natur.596..505P,Pleunis2021ApJ...911L...3P}), Medicina Northern Cross (MNC; \citealt{Trudu2022MNRAS.513.1858T}), Sardinia \citep{Pilia2020ApJ...896L..40P}, upgraded Giant Metrewave Radio Telescope (uGMRT; \citealt{Marthi2020MNRAS.499L..16M, Pleunis2021ApJ...911L...3P, Sand2022ApJ...932...98S}), Very Large Array (VLA; \citealt{Aggarwal2020ATel13664....1A,Aggarwal2020RNAAS...4...94A}).}

\tablenotetext{j}{Best localized with the Deep Synoptic Array (DSA), and the host spectroscopic redshift measured with Keck \citep{Ravi2022arXiv221109049R}.}

\tablenotetext{k}{Data sources: CHIME (CHIME/FRB VOEvent Service; \citealt{McKinven2022ATel15679....1M}), Arecibo \citep{Perera2022ATel15734....1P}, Big Scanning Antenna (BSA; \citealt{Fedorova2022ATel15713....1F}), DSA (\citealt{Ravi2022arXiv221109049R}; private communication), DSN \citep{Rajwade2022ATel15791....1R}, Lovell \citep{Rajwade2022ATel15791....1R}, MNC \citep{Pelliciari2022ATel15696....1P}, Stockert \citep{Kirsten2022ATel15727....1K,Herrmann2022ATel15691....1H}, Tianlai \citep{Yu2022ATel15758....1Y}.}

\tablenotetext{l}{Best localized with EVN \citep{Nimmo2022ApJ...927L...3N}, and the host spectroscopic redshift measured with MMT \citep{Fong2021ApJ...919L..23F}.}

\tablenotetext{m}{Data sources: CHIME (CHIME/FRB Public Database, \citealt{CHIME/FRB2021ATel14497....1C, Lanman2022ApJ...927...59L}), Allen Telescope Array (ATA; \citealt{Farah2021ATel14676....1F}), Apertif \citep{Atri2022ATel15197....1A}, Australian Square Kilometre Array Pathfinder (ASKAP; \citealt{Kumar2022MNRAS.512.3400K}), Effelsberg \citep{Hilmarsson2021MNRAS.508.5354H}, EVN \citep{Nimmo2022ApJ...927L...3N}, Five-hundred-meter Aperture Spherical radio Telescope (FAST; \citealt{Feng2022Sci...375.1266F, Wang2022ATel15288....1W, Xu2022Natur.609..685X, Zhang2022RAA....22l4002Z}), GBT \citep{Feng2022Sci...375.1266F}, Onsala \citep{Kirsten2021ATel14605....1K}, Parkes \citep{Kumar2022MNRAS.512.3400K}, Stockert \citep{Herrmann2021ATel14556....1H}, Usuda Deep Space Center (UDSC; \citealt{Takefuji2022ATel15285....1T,Ikebe2023PASJ...75..199I}), uGMRT \citep{Marthi2022MNRAS.509.2209M}, VLA \citep{Ravi2022MNRAS.513..982R}, Westerbork \citep{Ould-Boukattine2022ATel15190....1O, Ould-Boukattine2022ATel15192....1O}.}

\tablenotetext{n}{Best localized with EVN \citep{Marcote2017ApJ...834L...8M}, and the host spectroscopic redshift measured with Gemini-North \citep{Tendulkar2017ApJ...834L...7T}.}

\tablenotetext{o}{Data sources: Arecibo \citep{Spitler2014ApJ...790..101S,Scholz2016ApJ...833..177S,Spitler2016Natur.531..202S,Scholz2017ApJ...846...80S,MAGIC2018MNRAS.481.2479M,Michilli2018Natur.553..182M,Gourdji2019ApJ...877L..19G,Hessels2019ApJ...876L..23H,Aggarwal2021ApJ...922..115A,Hilmarsson2021ApJ...908L..10H,Hewitt2022MNRAS.515.3577H,Jahns2023MNRAS.519..666J}, Apertif \citep{Oostrum2017ATel10693....1O, Oostrum2020AA...635A..61O}, CHIME \citep{Josephy2019ApJ...882L..18J}, DSN \citep{Majid2020ApJ...897L...4M,Pearlman2020ApJ...905L..27P}, Effelsberg \citep{Hardy2017MNRAS.472.2800H,Spitler2018ApJ...863..150S,Houben2019AA...623A..42H,Cruces2021MNRAS.500..448C,Hilmarsson2021ApJ...908L..10H}, EVN \citep{Marcote2017ApJ...834L...8M}, FAST \citep{Li2021Natur.598..267L}, GBT \citep{Scholz2016ApJ...833..177S,Scholz2017ApJ...846...80S,Gajjar2018ApJ...863....2G,Michilli2018Natur.553..182M,Zhang2018ApJ...866..149Z,Hessels2019ApJ...876L..23H}, Lovell \citep{Rajwade2020MNRAS.495.3551R}, MeerKAT \citep{Caleb2020MNRAS.496.4565C}, VLA \citep{Chatterjee2017Natur.541...58C,Law2017ApJ...850...76L,Hilmarsson2021ApJ...908L..10H}.}

\tablenotetext{p}{Best localized with VLA, and the host spectroscopic redshift measured with the Hale Telescope  \citep{Niu2022Natur.606..873N}.}

\tablenotetext{q}{Data sources: FAST \citep{Niu2022Natur.606..873N}, GBT \citep{Anna-Thomas2022,Feng2022Sci...375.1266F}, Parkes \citep{Dai2022arXiv220308151D}, VLA \citep{Niu2022Natur.606..873N}.}

\tablenotetext{r}{Best localized with VLA, and the host spectroscopic redshift measured with Keck \citep{Bhandari2022AJ....163...69B}.}

\tablenotetext{s}{Data sources: Parkes \citep{Price2019MNRAS.486.3636P}, FAST \citep{Luo2020Natur.586..693L,Laha2022ApJ...930..172L}, VLA \citep{Bhandari2022AJ....163...69B}.}

\tablenotetext{t}{Localized with Chandra \citep{Israel2016MNRAS.457.3448I}.}

\tablenotetext{u}{Mean distance of a wide range estimated with various techniques ($\sim4.4$--$14.2$\,kpc; e.g., \citealt{Park2013ApJ...777...14P,Pavlovic2013ApJS..204....4P,Pavlovic2014SerAJ.189...25P,Surnis2016ApJ...826..184S,Kothes2018ApJ...852...54K,Mereghetti2020ApJ...898L..29M,Zhou2020ApJ...905...99Z}).}

\tablenotetext{v}{Extinction at the mean distance from the \cite{Green2019ApJ...887...93G} 3D dust map ($\sim5.2-7.2$\,mag over the whole distance range; caution that the dust map suffers a lack of bright main-sequence stars at the larger ditances), retrieved via \texttt{dustmaps} \citep{Green2018JOSS....3..695M}.}

\tablenotetext{w}{Data Sources: CHIME \citep{CHIME/FRB2020Natur.587...54C,Pleunis2020ATel14080....1P,Dong2022ATel15681....1D,Pearlman2022ATel15792....1P}, Survey for Transient Astronomical Radio Emission 2 (STARE2; \citealt{Bochenek2020Natur.587...59B}), FAST \citep{Zhang2020ATel13699....1Z}, GBT \citep{Maan2022ATel15697....1M}, Large Phased Array (LPA; \citealt{Rodin2022ARep...66...32R}), Yunnan \citep{Huang2022ATel15707....1H}, Wasterbork \citep{Kirsten2021NatAs...5..414K}.}

\end{deluxetable*}

We select a sample of FRBs that are well localized ($\lesssim2\arcsec$) and with known host-galaxy identifications,\footnote{See \url{https://frbhosts.org} for the up-to-date list of FRB host-galaxy identifications} located at decl.\,$\gtrsim-30^{\circ}$ in order to have access to the radio burst measurements from the Canadian Hydrogen Intensity Mapping Experiment/FRB \citep{CHIME/FRB2018ApJ...863...48C} Public Database\footnote{\url{https://www.chime-frb.ca/}} and optical forced photometry from the Zwicky Transient Facility (ZTF; \citealt{Bellm2019PASP..131a8002B, Graham2019PASP..131g8001G}) and Asteroid Terrestrial-impact Last Alert System (ATLAS; \citealt{Tonry2018PASP..130f4505T, Smith2020PASP..132h5002S})\footnote{Typical ZTF and ATLAS PSF FWHMs are $2\arcsec$ and $4\arcsec$, respectively, comparable to and larger than the adopted FRB localization error limit.} --- major FRB and optical time-domain surveys. We also include any other radio and optical data sets from the literature. We select both repeating and non-repeating FRBs; however, we find that all of the optical limits for non-repeating FRBs are not particularly constraining given the small number of bursts and large distances (see Appendix \ref{sec:norep}). Therefore, we focus on the repeating FRBs in the subsequent analysis. Their properties and references are summarized in Table~\ref{tab:FRBsample}. We note that the distance to the Galactic SGR~1935+2154 is not well constrained, which also results in a large extinction uncertainty.

\subsection{Optical Observations} \label{sec:opt}

For each FRB in our sample, we obtained optical photometry from the ZTF forced-photometry service\footnote{\url{https://ztfweb.ipac.caltech.edu/cgi-bin/requestForcedPhotometry.cgi}} \citep{Masci2019PASP..131a8003M} in the $g$, $r$, and $i$ bands, and the ATLAS forced photometry server\footnote{\url{https://fallingstar-data.com/forcedphot/}} \citep{Shingles2021TNSAN...7....1S} in the $c$ and $o$ bands for the entire time interval spanning all of the radio bursts for each FRB (i.e., from the first to latest measured bursts). The detection significance ($\sigma$) of optical measurements was determined from the ratio of measured flux ($f$) to its error ($f_{\rm err}$). For any measurements above $3\sigma$, we visually inspected the difference images and found that all were due to subtraction artifacts (e.g., bad focus, world coordinate system offset, shutter problems). Thus, we calculate the $3\sigma$ upper limits as $-2.5\,{\rm log}_{10}(3\times f_{\rm err})+{\rm ZP}$, where ``ZP" is the zero point in the AB magnitude system.

In addition, we carried out dedicated monitoring observations of the nearby FRB~20200120E in M81 \citep{Bhardwaj2021ApJ...910L..18B, Kirsten2022Natur.602..585K} with 300 s exposures simultaneously in the $g$, $r$, $i$, and $z$ bands from 2021 March 19 to 2022 May 18 (UT dates are used throughout) roughly every 10 days with MuSCAT3 \citep{Narita2020SPIE11447E..5KN} on the 2 m Faulkes Telescope North (Hawaii, USA) in the Las Cumbres Observatory (LCO; \citealt{Brown2013PASP..125.1031B}) network. These images have typical $griz$-band $3\sigma$ limiting magnitudes of $22.8$, $22.9$, $22.3$, and $22.0$, respectively.  The March 19 observations, during which no FRBs were announced, were used as template images, and image subtraction was performed using \texttt{lcogtsnpipe}\footnote{\url{https://github.com/LCOGT/lcogtsnpipe}} \citep{Valenti2016MNRAS.459.3939V}, a PyRAF-based photometric reduction pipeline, and \texttt{PyZOGY}\footnote{\url{https://github.com/dguevel/PyZOGY}} \citep{Guevel2017zndo...1043973G}, an implementation in Python of the subtraction algorithm described in \cite{Zackay2016ApJ...830...27Z}. The $griz$-band data were calibrated to AB magnitudes using the 13th Data Release of the Sloan Digital Sky Survey (SDSS; \citealt{Albareti2017ApJS..233...25A}).

\begin{figure*}
    \centering
    \gridline{\fig{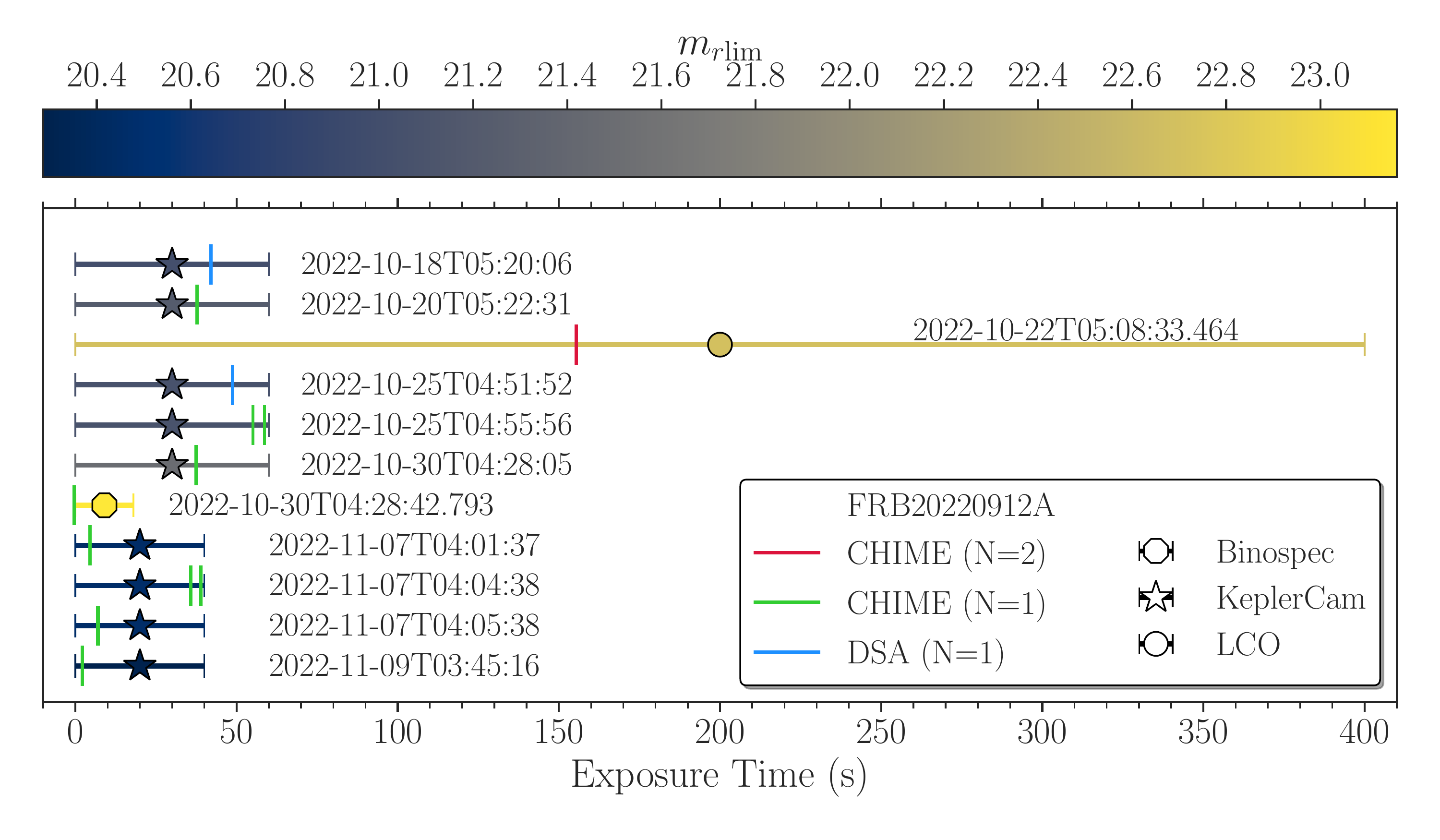}{0.84\textwidth}{}}
    \gridline{\fig{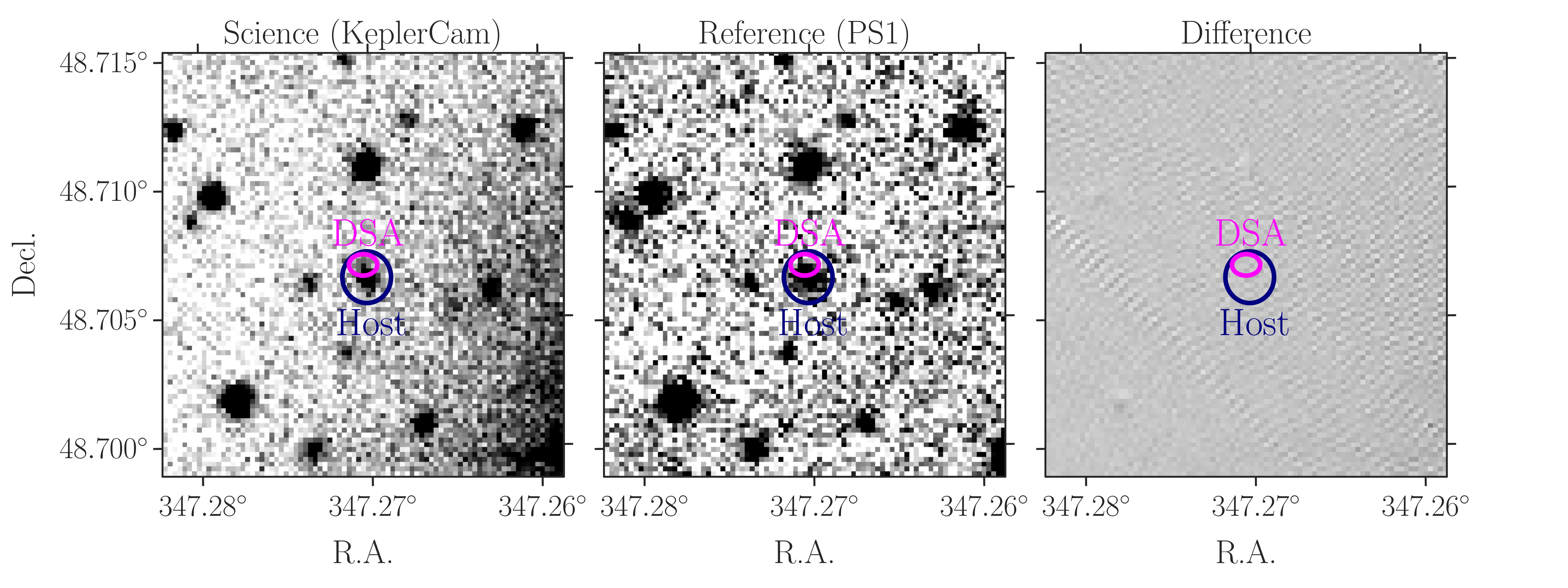}{0.99 \textwidth}{}}
\caption{Top: schematic timeline of the KeplerCam and LCO simultaneous exposures, as well as the temporally closest MMT/Binospec exposure with respect to the FRB~20220912A bursts detected by CHIME and DSA (where $N$ is the number of bursts detected within 1 s), color-coded by the observed $r$-band limiting magnitude. The UT date labeled next to each point is its exposure start time. In total, we covered 13 bursts in 10 simultaneous exposures: one DSA detection each in two KeplerCam exposures, one CHIME detection each in five KeplerCam exposures, and two CHIME detections each in two KeplerCam and one LCO exposures. 
Bottom: simultaneous KeplerCam 60 s image during the two CHIME detections (left), the corresponding PS1 template image (middle), and the resulting difference image (right), with the DSA localization region and host galaxy marked by the magenta ellipse ($\Delta{\rm R.A}\sim2\arcsec$ \& $\Delta{\rm decl.}\sim1\arcsec$; \citealt{Ravi2022arXiv221109049R}) and navy circle (Kron radius $= 3\farcs44$ from PS1 DR2; \citealt{Flewelling2020ApJS..251....7F}), respectively.  No optical counterpart is identified down to a $r$-band limiting magnitude of $21.1$.}
    \label{fig:simexp}
\end{figure*}

For the recently discovered highly active FRB~20220912A \citep{McKinven2022ATel15679....1M}, we carried out monitoring observations {\it during} the CHIME observing windows\footnote{\url{https://www.chime-frb.ca/astronomytools}} with three facilities: a series of 40 s exposures and a series of 60 s exposures in the $r$ band with KeplerCam \citep{Szentgyorgyi2005AAS...20711010S} on the 1.2 m Telescope at the Fred Lawrence Whipple Observatory (FLWO; Arizona, USA), reaching typical $3\sigma$ limiting magnitudes of $20.5$ and $21.1$, respectively; a series of 400 s exposures in the $r$ band with the Sinistro camera on the 1 m telescope at the McDonald Observatory (Texas, USA) in the LCO network, reaching a typical $3\sigma$ limiting magnitude of $22.6$; and a series of 15 s exposures in the $r$ band with Binospec \citep{Fabricant2019PASP..131g5004F} on the 6.5 m MMT Observatory (Arizona, USA), reaching a typical $3\sigma$ limiting magnitude of $23.2$. 
In total, we obtained 10 {\it simultaneous} optical observations during 13 FRB detections: nine with KeplerCam\footnote{One of which was reported in \cite{Hiramatsu2022ATel15699....1H}.} for the two Deep Synoptic Array (DSA) detections on 2022 October 18 and 25 \citep{Ravi2022arXiv221109049R} and the nine CHIME detections on October 20, 25, and 30 and November 7 and 9 (CHIME VOEvent Service\footnote{\url{https://www.chime-frb.ca/voevents}}); and one with LCO for the two CHIME detections on October 22. These KeplerCam and LCO simultaneous exposures, as well as a nearly simultaneous Binospec exposure 0.4 s after a CHIME detection, are shown in Figure~\ref{fig:simexp}. Due to the uncertainty ($\lesssim$ a few seconds) in the shutter opening time stamps (from the open command being issued to the shutter being fully opened), the nearly simultaneous Binospec exposure may indeed be simultaneous. We note this caveat in the following analysis and discussion wherever appropriate.

Using a custom photometry pipeline, the KeplerCam, LCO, and Binospec data were reduced and calibrated to AB magnitudes from the Pan-STARRS1\footnote{The field is not covered by SDSS.} (PS1; \citealt{Chambers2016arXiv161205560C}) Data Release 2 (DR2; \citealt{Flewelling2020ApJS..251....7F}). Cosmic rays were identified and masked using \texttt{Cosimic-CONN}\footnote{\url{https://github.com/cy-xu/cosmic-conn}} \citep{Xu2021arXiv210614922X,xu_chengyuan_2021_5034763,Xu2022arXiv220410356X}, and image subtraction was performed against PS1 template images using \texttt{PyZOGY}. For each simultaneous exposure, we stacked it with the subsequent exposures in the same series to obtain deeper limits (typical limiting magnitudes of $21.5$ and $23$ for KeplerCam and LCO observations, respectively). An example image subtraction is shown in Figure~\ref{fig:simexp} for one of the simultaneous KeplerCam images. We do not detect any transient source in the subtracted KeplerCam, LCO, and Binospec images within any of the FRB localization regions. Thus, we report their $3\sigma$ upper limits. 

We correct all optical limits for the MW extinction assuming the \citet{Fitzpatrick1999PASP..111...63F} reddening law with $R_V=3.1$. After correcting the DM- and frequency-dependent time delay (see Equation (1) of \citealt{Cordes2019ARA&A..57..417C}) and referencing all the radio and optical measurements to the solar system barycenter \citep{Eastman2010PASP..122..935E},\footnote{\url{https://github.com/tronsgaard/barycorr}} we find the temporally closest radio burst for each optical limit with respect to the optical exposure midpoint to determine the time difference.

\section{Analysis and Discussion} 
\label{sec:ana}

\subsection{Optical Luminosity Limits} \label{sec:luminosity}

\begin{figure*}
    \centering
    \includegraphics[width=0.925\textwidth]{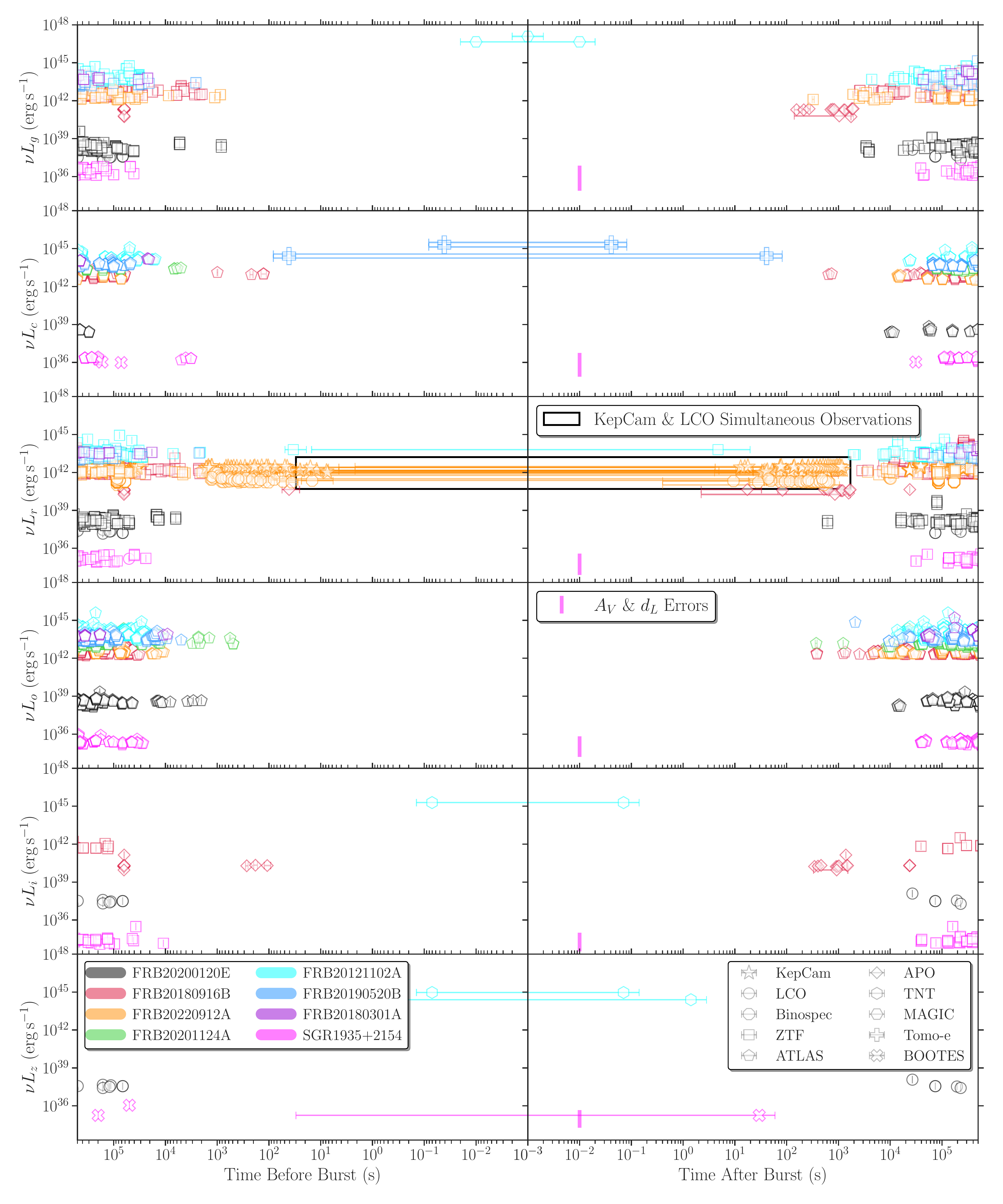}
\caption{Optical luminosity limits (in the $g$, $c$, $r$, $o$, $i$, and $z$ bands from the top to bottom rows) with respect to the temporally closest FRB detection for our sample of eight repeating FRBs (including the Galactic SGR~1935+2154). Representative limits are shown for the targeted high-speed observations (TNT $i$ and $i+z$, MAGIC $U$, Tomo-e open) in the panels with the closest filter effective wavelength, and similarly for LCO $R$ and BOOTES $Z$ and clear bands. A typical luminosity error in each band for SGR~1935+2154 (due to the large extinction and distance uncertainties) is shown as the magenta vertical bar. Our KeplerCam and LCO limits for FRB~20220912A (enclosed by the black rectangle) are the deepest among the \textit{simultaneous} observations of extragalactic FRBs. Optical data sources: KeplerCam (this work for FRB~20220912A), LCO (this work and \citealt{Lin2020Natur.587...63L} for FRBs~20200120E and 20220912A in $griz$, and SGR~1935+2154 in $R$, respectively), Binospec (this work for FRB~20220912A), ZTF (\citealt{Andreoni2021ATel14666....1A, Andreoni2020ApJ...896L...2A} for FRBs~20200120E and 20180916B, respectively), ZTF forced-photometry service (\citealt{Masci2019PASP..131a8003M} and this work for all the FRBs), ATLAS forced photometry server (\citealt{Shingles2021TNSAN...7....1S} and this work for all the FRBs), APO (\citealt{Kilpatrick2021ApJ...907L...3K} for FRB~20180916B), TNT (\citealt{Hardy2017MNRAS.472.2800H} for FRB~20121102A), MAGIC (\citealt{MAGIC2018MNRAS.481.2479M} for FRB~20121102A), Tomo-e (\citealt{Niino2022ApJ...931..109N} for FRB~20190520B), and BOOTES (\citealt{Lin2020Natur.587...63L} for SGR~1935+2154). (The data used to create this figure are available.)}
    \label{fig:lumlim}
\end{figure*}

None of our simultaneous and follow-up observations or the archival observations have led to a detected optical counterpart. We plot all of the resulting KeplerCam, LCO, and Binospec luminosity limits, along with the archival and literature sample in Figure~\ref{fig:lumlim}. Most of the untargeted ZTF and ATLAS observations occur $\gtrsim 10^{4}$ s before and/or after radio bursts, with a wide luminosity limit range of $\sim 10^{35}$--$10^{45}$\,erg\,s$^{-1}$; the deepest limits are for the Galactic magnetar SGR\,1935+2154, given the much smaller distance compared to the other extragalactic FRBs. An untargeted ZTF $r$-band observation for FRB~20200120E \citep{Andreoni2021ATel14666....1A} and the targeted (but nonsimultaneous) $gri$-band observations with ARCTIC on the Apache Point Observatory (APO) 3.5 m telescope \citep{Huehnerhoff2016SPIE.9908E..5HH} for FRB~20180916B \citep{Kilpatrick2021ApJ...907L...3K} probe a luminosity range of $\sim 2\times10^{38}-4\times10^{40}$\,erg\,s$^{-1}$ within $\sim 10^3$ s of radio bursts.

On a shorter timescale, several targeted high-speed simultaneous observations --- by ULTRASPEC on the 2.4 m Thai National Telescope (TNT; \citealt{Tulloch2011MNRAS.411..211T,Dhillon2014MNRAS.444.4009D}) for FRB~20121102A \citep{Hardy2017MNRAS.472.2800H}; the central pixel of the Major Atmospheric Gamma Imaging Cherenkov (MAGIC) telescopes \citep{Lucarelli2008NIMPA.589..415L,Hassan2017ICRC...35..807H} for FRB~20121102 \citep{MAGIC2018MNRAS.481.2479M}; and Tomo-e Gozen on the Kiso 105 cm Schmidt telescope \citep{Sako2018SPIE10702E..0JS} for FRB~20190520B \citep{Niino2022ApJ...931..109N} --- span a temporal range $\sim 1$--$10$ ms before/after radio bursts with a luminosity upper limit range of $\sim 10^{45}$--$10^{47}$\,erg\,s$^{-1}$ in a single exposure (longer and deeper stacked exposures are also shown in Figure~\ref{fig:lumlim}). In the ZTF archival search, we also find a simultaneous $30$ s $r$-band observation of FRB~20121102A with a luminosity limit of $\sim 6\times10^{43}$\,erg\,s$^{-1}$.

Among the simultaneous optical observations, our KeplerCam and LCO $r$-band limits for FRB~20220912A are the deepest to date in terms of luminosity for the extragalactic FRBs, $\sim (0.3$--$2.9)\times10^{42}$\,erg\,s$^{-1}$ with $30$--$400$ s exposures (i.e., excluding the Burst Observer and Optical Transient Exploring System, BOOTES, $Z$-band limit of $\sim 1.8\times10^{35}$\,erg\,s$^{-1}$ with a 60 s exposure for the Galactic SGR~1935+2154; \citealt{Lin2020Natur.587...63L}). Our nearly simultaneous Binospec limit for FRB~20220912A is at $\sim 2\times 10^{41}$\,erg\,s$^{-1}$ with a $15$ s exposure started at $0.4$ s after a burst.

\subsection{Optical-to-Radio Flux and Fluence Ratio Limits} 
\label{sec:flux}

\begin{figure*}
    \centering
    \includegraphics[width=0.925\textwidth]{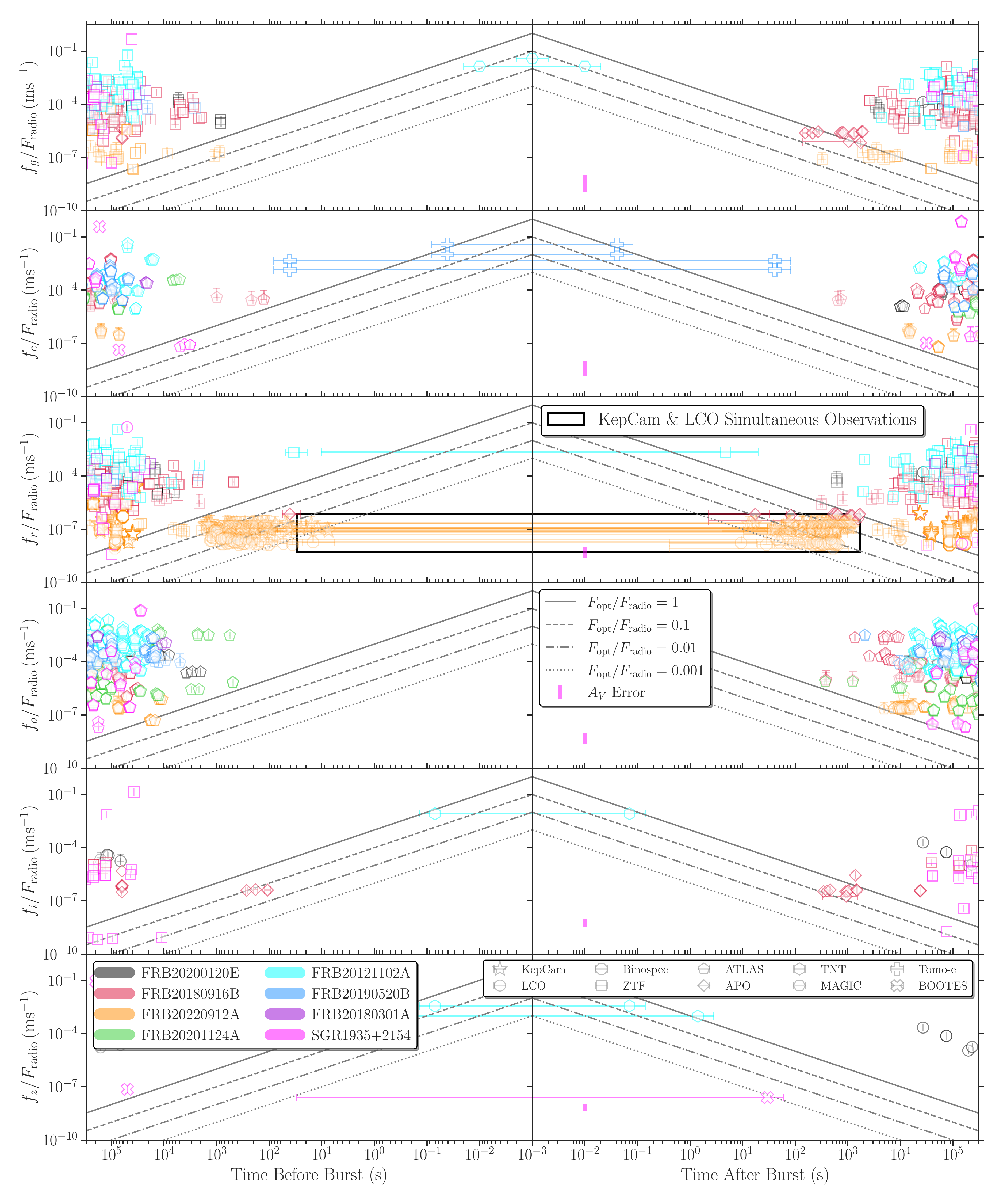}
\caption{Ratio of optical flux limit (in janskys; in the $g$, $c$, $r$, $o$, $i$, and $z$ bands from the top to bottom rows) to radio fluence (in janskys times millisecond) with respect to the temporally closest FRB detection for our sample of eight repeating FRBs
(including the Galactic SGR~1935+2154); labels are as in Figure~\ref{fig:lumlim}. In the case when the radio fluence of a particular burst is not available, it is assumed to be the mean of the fluence distribution of that particular repeating FRB (and shown as a transparent point with an upper error bar corresponding to one standard deviation). The dashed and dotted gray lines in each panel show $F_{\rm opt}/F_{\rm radio}=1$, $0.1$, $0.01$, and $0.001$ assuming $F_{\rm opt}=f_{\rm opt}\times \Delta t$, where $\Delta t$ is the time delay between the optical and radio observations. Note that most of the {\it simultaneous} observations (KeplerCam, LCO, TNT, MAGIC, Tome-e, BOOTES), targeted Binospec and APO, and some untargeted ZTF observations are in the regime of $F_{\rm opt}/F_{\rm radio}\lesssim 1$. Our KeplerCam and LCO limits for FRB~20220912A are comparable even to the BOOTES simultaneous observation of SGR~1935+2154 ($f_{\rm opt}/F_{\rm radio}\lesssim 10^{-7}$\,ms$^{-1}$, or $F_{\rm opt}/F_{\rm radio}\lesssim 10^{-3}$). Our nearly simultaneous Binospec limit is at $f_{\rm opt}/F_{\rm radio}\lesssim 10^{-8}$\,ms$^{-1}$, or $F_{\rm opt}/F_{\rm radio}\lesssim 10^{-4}$ at $\Delta t\approx0.4$\,s.
    }
    \label{fig:ftoF}
\end{figure*}

To place the optical limits in the context of the individual FRB burst properties, we plot the ratio of optical flux limit to FRB radio fluence, $f_{\rm opt}/F_{\rm radio}$, in Figure~\ref{fig:ftoF}. We make this parameter choice because the optical burst duration ($t_{\rm opt}$) is not known. If the fluence measurement of a particular radio burst is not yet published, it is assumed to be the mean of the fluence distribution of the relevant repeating FRB (e.g., the recently discovered FRB~20220912A). We find that the range of flux-to-fluence ratio limits is $\sim 10^{-9}$--$0.1$ ms$^{-1}$, on a timescale of $\gtrsim 10^3$ s.

If we make a conservative assumption that the optical counterpart duration is $t_{\rm opt}\approx \Delta t$, where $\Delta t$ is the time delay between an FRB and an optical observation, we find that all of the targeted simultaneous observations lie below the critical line of $f_{\rm opt}\Delta t/F_{\rm radio}\approx F_{\rm opt}/F_{\rm radio} = 1$.  On the other hand, nearly all of the untargeted and archival observations are well above the $F_{\rm opt}/F_{\rm radio} = 1$ line (Figure~\ref{fig:ftoF}), meaning that they provide weak constraints for any model in which the energy release per frequency in the optical band is at most comparable to that in the radio bursts; exceptions are some nonsimultaneous APO and untargeted ZTF observations for FRBs~20180916 and 20220912A, respectively. Our nearly simultaneous Binospec limit is at $f_{\rm opt}/F_{\rm radio}\lesssim 10^{-8}$\,ms$^{-1}$ (or $F_{\rm opt}/F_{\rm radio}\lesssim 10^{-4}$) at $\Delta t\approx0.4$\,s, which is the deepest for any FRBs on this timescale.

Among the simultaneous optical observations, our KeplerCam and LCO $r$-band limits for FRB~20220912A are the deepest to date in terms of $f_{\rm opt}/F_{\rm radio}$ for any extragalactic FRB, and indeed comparable to limits for the Galactic SGR~1935+2154, with $f_{\rm opt}/F_{\rm radio}\lesssim 10^{-7}$ ms$^{-1}$. In terms of fluence ratios, our observations place a limit of $F_{\rm opt}/F_{\rm radio}\lesssim 10^{-3}$. Our nearly simultaneous Binospec limit would be the deepest if it were indeed simultaneous given the aforementioned shutter timing uncertainty.
We stress that these fluence ratio limits are lower than the observed values for the Crab and Geminga pulsars (e.g., \citealt{Danilenko2011MNRAS.415..867D,Buhler2014RPPh...77f6901B}), which may suggest a different emission mechanism (although pulsars with lower fluence ratios have also been observed; see, e.g., \citealt{Niino2022ApJ...931..109N} for a discussion for FRB~20190520B in this context).

\subsection{Constraints on Fast Optical Burst Models} \label{sec:fob}

\begin{figure*}
    \centering
    \includegraphics[width=0.95\textwidth]{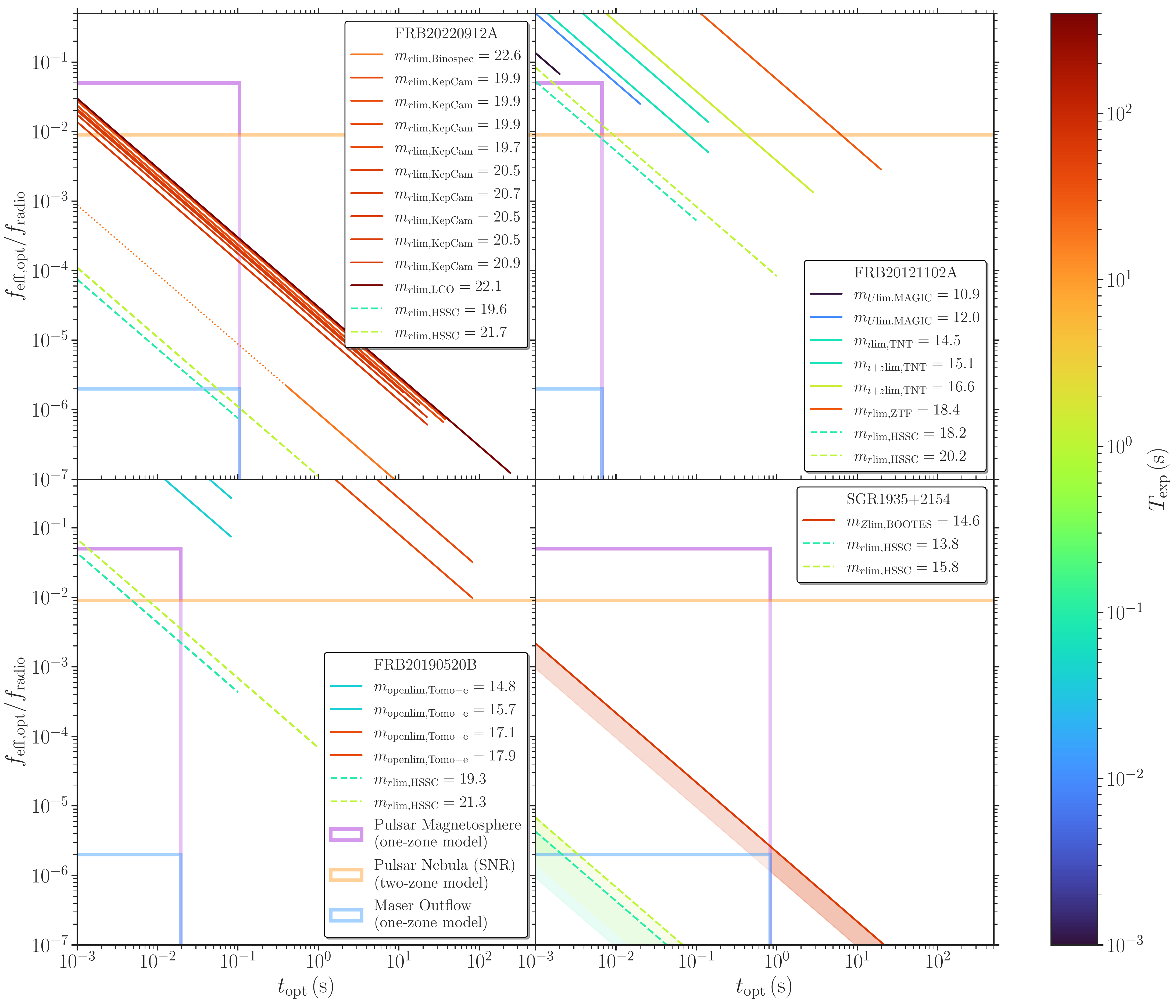}
\caption{Effective optical flux limit ($f_{\rm eff,opt}=f_{\rm opt}T_{\rm exp}/t_{\rm opt}$) to radio flux ratio with respect to the optical burst duration ($t_{\rm opt}$), color-coded by the exposure time ($T_{\rm exp}$), for the sample of FRBs with simultaneous optical observations (including the nearly simultaneous Binospec observation for FRB~20220912A). The quoted magnitude limits are the extinction-corrected values for each FRB, and their solid lines extend from the time of radio burst to the end of each exposure. The dotted extension to the nearly simultaneous Binospec observation shows $f_{\rm eff,opt}$ within the timing uncertainty. The shaded regions show the flux error ranges of the Galactic SGR~1935+2154 due to the large extinction uncertainty.
Broad ranges of optical burst model predictions from \citet{Yang2019ApJ...878...89Y} are also plotted as rectangular regions, where the duration of one-zone models are shown with a width corresponding to the mean plus one standard deviation of the $t_{\rm FRB}$ distribution of each FRB. Potential constraints from future observations with Subaru HSSC are also plotted as dashed lines for each FRB. Note that the optical limits of FRBs~20121102A and 20190520B barely reach the pulsar nebula model parameter space, while our limits for FRB~20200912A and published limit for SGR~1935+2154 reach the pulsar magnetosphere as well as the pulsar nebula model parameter spaces.
    }
    \label{fig:ftofmod}
\end{figure*}

To constrain possible fast optical bursts on a comparable timescale to the radio bursts, we define an effective optical flux limit on a timescale, $t_{\rm opt}$ (see also \citealt{Lyutikov2016ApJ...824L..18L}): 
\begin{equation}
f_{\rm eff, opt} = f_{\rm opt} \frac{T_{\rm exp}}{t_{\rm opt}},
\label{eq:feff}
\end{equation}
where $T_{\rm exp}$ is the exposure time. For the sample of FRBs with simultaneous optical observations, we show $f_{\rm eff,opt}/f_{\rm radio}$ as a function of $t_{\rm opt}$ in Figure~\ref{fig:ftofmod}. The previously published high-speed observations of FRBs~20121102A and 20190520B only probe $f_{\rm eff, opt}/f_{\rm radio}\gtrsim 0.1$ on millisecond timescales, and reach $f_{\rm eff, opt}/f_{\rm radio}\sim 0.01$ only on $\gtrsim {\rm second}$ timescales due to the shallow integrated flux limit (i.e., apparent magnitude limit).

On the other hand, our KeplerCam and LCO observations for FRB~20220912A are far more sensitive, reaching $f_{\rm eff,opt}/f_{\rm radio}\lesssim 0.02$ on a milliseocnd timescale, and $\lesssim 3\times 10^{-5}$ on a second timescale.  These limits are only an order of magnitude higher than the limits for the Galactic SGR~1935+2154. Our nearly simultaneous Binospec observation for FRB~20220912A reaches $f_{\rm eff,opt}/f_{\rm radio}\lesssim 10^{-6}$ on a second timescale (and $\lesssim10^{-3}$ on a millisecond timescale if it were indeed simultaneous), which is on the same order as the SGR~1935+2154 limits.

We also plot in Figure~\ref{fig:ftofmod} broad ranges of some optical burst models from \cite{Yang2019ApJ...878...89Y}.\footnote{Optical emission models from the intrinsic FRB mechanisms are not shown here since they are expected to have much lower $f_{\rm opt}/f_{\rm radio}$ (e.g., $<10^{-9}$ and $<10^{-10}$ from the maser emission and curvature radiation models, respectively; \citealt{Kumar2017MNRAS.468.2726K,Waxman2017ApJ...842...34W,Lu2018MNRAS.477.2470L,Lu2020MNRAS.498.1397L}).} In these models the subsequent optical bursts to FRBs originate from several inverse-Compton scattering processes involving a highly magnetized central source, such as a young pulsar or magnetar. One-zone models (e.g., pulsar magnetosphere and maser outflow) are expected to have $t_{\rm opt}\approx t_{\rm FRB}$, while two-zone models could result in a much longer timescale for the optical emission given the larger scattering region than FRB emission region (e.g., $\sim$\,$5000$\,s for the pulsar nebula model; \citealt{Yang2019ApJ...878...89Y}). The optical constraints for FRBs~20121102A and 20190520B from previous observations barely reach the upper end of the pulsar nebula model with $t_{\rm opt}\sim 0.1$--$10$\,s. Those of SGR~1935+2154 are well within the model predictions for pulsar magnetosphere and nebula (see also \citealt{Lin2020Natur.587...63L}). For the first time for an extragalactic FRB, our limits for FRB~20220912A also reach the model prediction ranges.  Within these models, these limits provide meaningful constraints on the magnetic field of $B\lesssim10^{14}$\,G for magnetars or the spin period of $P\lesssim0.01$\,s for young pulsars.

In Figure~\ref{fig:ftofmod}, we also show potential constraints from high-speed observations with a large-aperture telescope, specifically the Subaru High-Speed Suprime-Cam (HSSC), a planned upgrade for the Suprime-Cam instrument. These potential future observations can provide limits of $f_{\rm eff, opt}/f_{\rm radio}\lesssim0.01$--$10^{-4}$ on millisecond to second timescales for FRBs~20121102A and 20190520B, reaching the pulsar magnetosphere and pulsar nebula model parameter space. For FRB~20220912A and SGR~1935+2154, these observations would probe $f_{\rm eff, opt}/f_{\rm radio} \lesssim10^{-4}$--$10^{-5}$ on a millisecond timescale to $10^{-7}$--$10^{-8}$ on a second timescale, reaching even to the maser outflow model parameter space.

\subsection{Constraints on the Synchrotron Maser Model} \label{sec:maser}

To place the luminosity limits on a longer timescale into context, we compare them with theoretical light curves from the synchrotron maser model \citep{Metzger2019MNRAS.485.4091M,Margalit2020ApJ...899L..27M, Margalit2020MNRAS.494.4627M}. In this model, relativistic plasma ejected from a central engine (e.g., a magnetar) creates a shock in the external magnetized medium (produced in previous flare events), resulting in an FRB from synchrotron maser emission \citep{Plotnikov2019MNRAS.485.3816P}. The model predicts that a broadband synchrotron afterglow will follow the FRB and peak in gamma- to X-rays on a millisecond timescale and in optical on minute to hour timescales. We provide details of the model (including the relevant equations) in Appendix~\ref{sec:mod} and show model light curves in Figure~\ref{fig:mod}. Assuming an external medium with a constant (shell-like) density distribution, the afterglow light curves are parameterized by the flare energy ($E_{\rm flare}$) ejecting the relativistic plasma and the external density ($n_{\rm ext}$) at the shock front; we note that the density structure could in reality be more complex, either wind-like ($\propto r^{-2}$) or a combination of multiple structures (e.g., see \citealt{Cooper2022MNRAS.517.5483C} for the case of SGR~1935+2154).

\begin{figure*}
    \centering
    \includegraphics[width=0.95\textwidth]{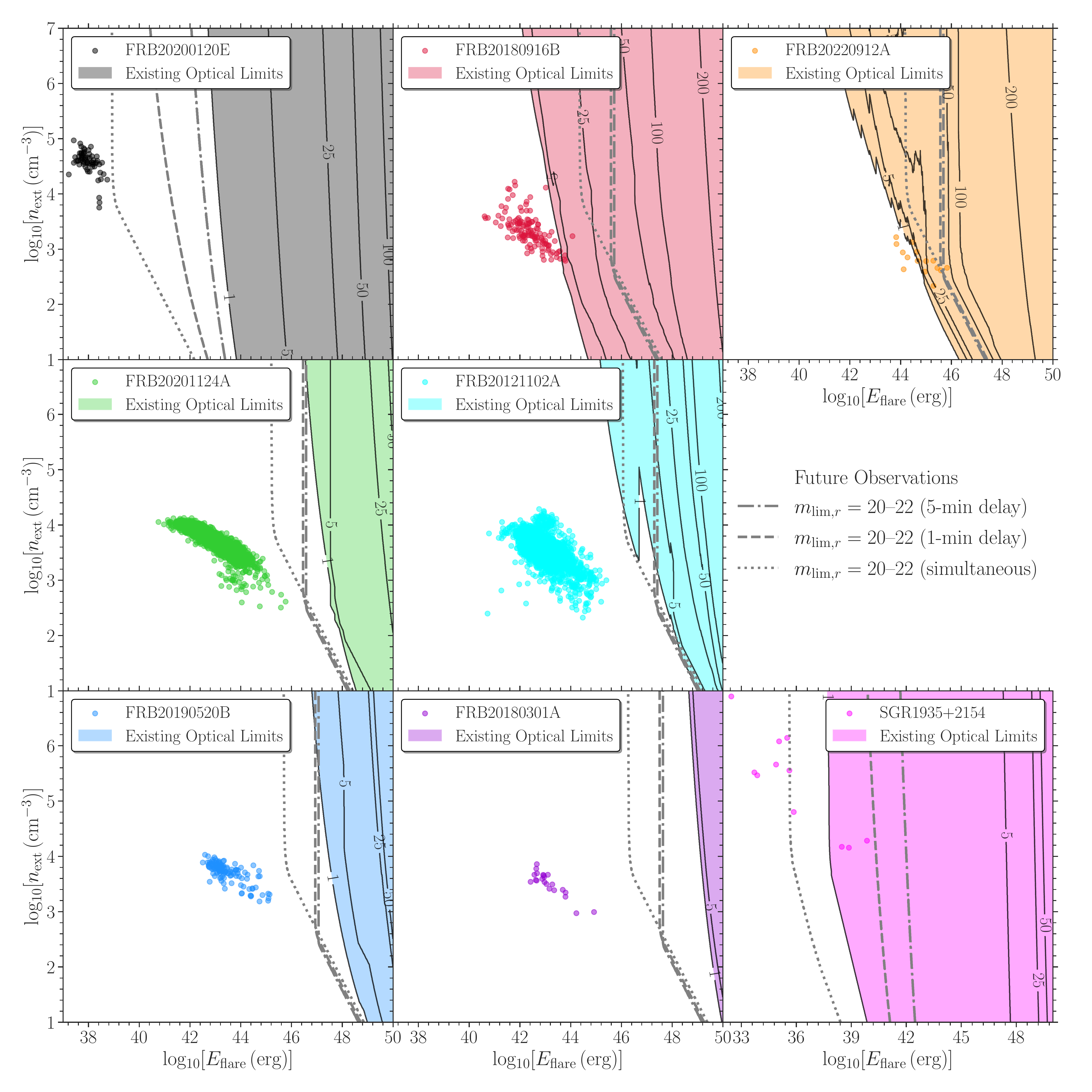}
\caption{FRB-based estimates (points) and optical-based constraints (shaded regions) on the external density ($n_{\rm ext}$) and flare energy ($E_{\rm flare}$) in the synchrotron maser model (Figure~\ref{fig:mod}). The contours mark the number of most stringent optical limits (i.e., deeper limits with shorter delays; from left to right in each panel: 1, 5, 25, 50, 100, and 200) used to constrain the parameter space, while the gray dashed and dotted lines show the possible constraints from future observations: $60$--$300$ s exposures (limiting magnitude of $20$--$22$) with $1$--$5$ minutes delay using a 1 m class telescope (e.g., KeplerCam and LCO), and simultaneous $0.1$--$1$ s exposure (limiting magnitude of $20$--$22$) using an 8 m class telescope (e.g., Subaru HSSC). Note that optical limits already reach the high-$E_{\rm flare}$ ends of the distributions for FRB~20180916B and SGR~1935+2154, and down to the mid-$E_{\rm flare}$ range ($\approx65\%$) of the distribution of FRB~20220912A with our KeplerCam, LCO, and Binospec observations.
The future observations have potential to reach FRBs~20200120E and 20201124A, and to constrain SGR~1935+2154 even further.
    }
    \label{fig:const}
\end{figure*}

We explore $E_{\rm flare}$ and $n_{\rm ext}$ ranges of $10^{37}$--$10^{50}$\,erg and $10$--$10^{7}$\,cm$^{-3}$, respectively, for each FRB, except the Galactic SGR~1935+2154, where we use a lower $E_{\rm flare}$ bound of $10^{32}$\,erg. We calculate a model light curve (in terms of specific luminosity at each optical filter effective wavelength) for every combination of $E_{\rm flare}$ and $n_{\rm ext}$ and compare each optical limit with the integrated average over the exposure time. If the model light curve is brighter than the optical limit, we consider the specific combination of $E_{\rm flare}$ and $n_{\rm ext}$ parameter space as constrained. The results are shown in Figure~\ref{fig:const}, along with FRB-based estimates of $E_{\rm flare}$ and $n_{\rm ext}$ for each burst (see Equations~(\ref{eq:Eradio})--(\ref{eq:next})). The model light curves, and thus the optical constraints, are more sensitive to $E_{\rm flare}$ than to $n_{\rm ext}$ since the characteristic synchrotron frequency, above which the light curves decline exponentially, is set by $E_{\rm flare}$ (see Equations~(\ref{eq:nusync0})--(\ref{eq:nuLnu})). 

As shown in Figure~\ref{fig:const}, the existing optical constraints already reach the high-$E_{\rm flare}$ ends of the distributions for FRB~20180916B ($\gtrsim$\,$10^{44}$\,erg; see also \citealt{Kilpatrick2021ApJ...907L...3K}) and SGR~1935+2154 ($\gtrsim$\,$10^{39}$\,erg; see also \citealt{Cooper2022MNRAS.517.5483C}). The combination of our simultaneous KeplerCam and LCO observations, and nearly simultaneous deep Binospec observation (with only a $0.4$ s delay), are the main drivers for constraining the FRB~20220912A parameter space ($\gtrsim$\,$10^{45}$\,erg) covering $\approx65\%$ of its radio bursts observed to date (the ones with published flux/fluence information).

In Figure~\ref{fig:const}, we also show the potential constraints from future observations.
Follow-up $60$-$300$ s observations with $1$--$5$ minutes delay using a 1 m class telescope (e.g., KeplerCam and LCO; limiting magnitude of $20$--$22$) have the potential to reach the high-$E_{\rm flare}$ end of the distribution of FRB~20201124A ($\sim$\,$10^{46}$\,erg).
Similarly, simultaneous $0.1$--$1$ sec observations using an 8 m class telescope (e.g., Subaru HSSC) could potentially constrain the high-$E_{\rm flare}$ end and mid-$E_{\rm flare}$ range of the distributions of FRB~20200120E ($\sim$\,$10^{39}$\,erg) and SGR~1935+2154 ($\sim$\,$10^{36}$\,erg), respectively, if we could target them in an active phase.

\section{Summary and Conclusions} 
\label{sec:sum}

We presented the most constraining limits on transient optical counterparts of FRBs to date, from dedicated {\it simultaneous} observations of 13 bursts from the repeating FRB~20220912A, and a {\it nearly simultaneous} observation with a 6.5 m MMT/Binospec delayed by only 0.4 s (which may indeed be simultaneous given the shutter timing uncertainty). We further presented our dedicated follow-up observations, and compiled all available serendipitous archival and published observations following bursts from a sample of eight repeating FRBs (including the Galactic SGR~1935+2154) and seven non-repeating FRBs. This data set, and in particular our simultaneous observations of FRB~20220912A, provides the largest study to date of optical emission associated with FRBs.  None of the FRBs studied here presented detectable transient optical counterparts. Our key findings are as follows:

\vspace{-5pt}
\begin{enumerate}
  \item Nearly all archival optical observations have a delay of $\gtrsim 10^4$ s relative to the time of bursts, while targeted observations have shorter and even up to no delays providing some constraining limits.  
\vspace{-5pt}
  \item Previous simultaneous observations provided optical luminosity limits of $\lesssim 10^{45}$\,erg\,s$^{-1}$ (with the exception of SGR~1935+2154 with $\lesssim 10^{35}$\,erg\,s$^{-1}$), while our simultaneous KeplerCam and LCO observations provide limits of $\lesssim 10^{42}$\,erg\,s$^{-1}$, a three orders of magnitude improvement. Our nearly simultaneous Binospec observation reaches a limit of $\lesssim 2\times10^{41}$\,erg\,s$^{-1}$.
\vspace{-5pt}
  \item Normalizing the optical fluence limits (using the time delay as a proxy for $t_{\rm opt}$) by the radio fluences, most archival observations have $F_{\rm opt}/F_{\rm radio}\gtrsim 1$, and therefore place only weak constraints on the relative energy per frequency in the optical and radio. On the other hand, our simultaneous KeplerCam and LCO observations and nearly simultaneous deep Binospec observations place more meaningful limits of $F_{\rm opt}/F_{\rm radio}\lesssim 10^{-3}$ and $\lesssim 10^{-4}$, respectively (even comparable to/deeper than the limits obtained for SGR~1935+2154).
\vspace{-5pt}
  \item Comparing the simultaneous limits (normalized by radio flux), we find that previous observations did not constrain models for simultaneous optical bursts (with the exception of one observation of SGR~1935+2154), while our KeplerCam and LCO and nearly simultaneous Binospec observations of FRB~20220912A place the first meaningful constraints on model predictions, ruling out portions of the parameter space for some models.
\vspace{-5pt}
  \item Interpreting the optical luminosity limits in the context of the synchrotron maser model, and comparing to the distributions of $E_{\rm flare}$ and $n_{\rm ext}$ inferred from the radio bursts for each repeating FRB, we find that most optical limits are not constraining (except the one limit for SGR~1935+2154), but our simultaneous and targeted observations of FRB~20220912A rule out the bulk of the parameter space for this model.
\end{enumerate}

Given the estimated high burst rate of FRB~20220912A ($\sim$ a few hundreds per hour; \citealt{Kirsten2022ATel15727....1K,Bhusare2022ATel15806....1B,Feng2022ATel15723....1F,McKinven2022ATel15679....1M,Zhang2022ATel15733....1Z}), and the fact that most searches have not published arrival times for their detected bursts, it is possible that we have simultaneous coverage for additional bursts (including multiple bursts in a single exposure added coherently, as in the KeplerCam and LCO exposures), which would place tighter constraints on the fast optical burst and synchrotron maser models. This will be explored once all the FRB measurements (e.g., arrival times, duration, flux, and fluence) become public.  Indeed, to enable this type of work on a rapid timescale, we advocate for FRB search efforts to make these basic properties public in real time.

We further investigate the potential of future follow-up and coordinated simultaneous observations using $1$--$8$ m class telescopes on a wide range of timescales from milliseconds to minutes. We find that these observations have the potential to better constrain optical emission from FRBs compared to the bulk of serendipitous archival observations.  In particular, based on the experience and constraining limits from our KeplerCam, LCO, and Binospec observations, as well as previous targeted observations (e.g, \citealt{Hardy2017MNRAS.472.2800H,MAGIC2018MNRAS.481.2479M,Kilpatrick2021ApJ...907L...3K,Niino2022ApJ...931..109N}), we advocate for an approach of {\it simultaneous monitoring} of FRBs during CHIME (or other facility) observing windows, especially when repeating FRBs are in a particularly active phase.

\acknowledgments

We are grateful to Charles Kilpatrick, Wenbin Lu, Lachlan Marnoch, Sandro Mereghetti, Eli Waxman, and Luca Zampieri for useful discussions and comments, to Joscha Jahns for providing Arecibo burst information of FRB~20121102A, to Vikram Ravi for providing DSA burst information of FRB~20220912A, to Daniel Fabricant, Jan Kansky, and Benjamin Weiner for scheduling the MMT Binospec observations and providing the instrument details, and to Griffin Hosseinzadeh for providing the basis of our custom photometry pipeline.  

The Berger Time-Domain research group at Harvard is supported by the NSF and NASA.
I.A. is a CIFAR Azrieli Global Scholar in the Gravity and the Extreme Universe Program and acknowledges support from that program, from the European Research Council (ERC) under the European Union’s Horizon 2020 research and innovation program (grant agreement No. 852097), from the Israel Science Foundation (grant No. 2752/19), from the United States--Israel Binational Science Foundation (BSF), and from the Israeli Council for Higher Education Alon Fellowship.

This work makes use of observations from KeplerCam on the 1.2 m telescope at the Fred Lawrence Whipple Observatory. Observations reported here were obtained at the MMT Observatory, a joint facility of the Smithsonian Institution and the University of Arizona.

This work makes use of observations from the Las Cumbres Observatory global telescope network. This paper is based in part on observations made with the MuSCAT3 instrument, developed by the Astrobiology Center and under financial support by JSPS KAKENHI (grant No. JP18H05439) and JST PRESTO (grant No. JPMJPR1775), at Faulkes Telescope North on Maui, HI, operated by the Las Cumbres Observatory. The authors wish to recognize and acknowledge the very significant cultural role and reverence that the summit of Haleakal$\bar{\text{a}}$ has always had within the indigenous Hawaiian community. We are most fortunate to have the opportunity to conduct observations from the mountain. 

This research has made use of the NASA Astrophysics Data System (ADS), the NASA/IPAC Extragalactic Database (NED), and NASA/IPAC Infrared Science Archive (IRSA, which is funded by NASA and operated by the California Institute of Technology), and IRAF (which is distributed by the National Optical Astronomy Observatory, NOAO, operated by the Association of Universities for Research in Astronomy, AURA, Inc., under cooperative agreement with the NSF).

We acknowledge use of the CHIME/FRB Public Database and the VOEvent Service, provided at \url{https://www.chime-frb.ca/} by the CHIME/FRB Collaboration.




This work has made use of data from the Zwicky Transient Facility (ZTF).  ZTF is supported by NSF grant No. AST-1440341 and a collaboration including Caltech, IPAC, the Weizmann Institute for Science, the Oskar Klein Center at Stockholm University, the University of Maryland, the University of Washington, Deutsches Elektronen-Synchrotron and Humboldt University, Los Alamos National Laboratories, the TANGO Consortium of Taiwan, the University of Wisconsin at Milwaukee, and Lawrence Berkeley National Laboratories. Operations are conducted by COO, IPAC, and UW. The ZTF forced-photometry service was funded under the Heising-Simons Foundation grant
No. 12540303 (PI: Graham).

This work has made use of data from the Asteroid Terrestrial-impact Last Alert System (ATLAS) project. ATLAS is primarily funded to search for near-Earth asteroids through NASA grant Nos. NN12AR55G, 80NSSC18K0284, and 80NSSC18K1575; byproducts of the NEO search include images and catalogs from the survey area. This work was partially funded by Kepler/K2 grant No. J1944/80NSSC19K0112 and HST grant No. GO-15889, and STFC grant Nos. ST/T000198/1 and ST/S006109/1. The ATLAS science products have been made possible through the contributions of the University of Hawaii Institute for Astronomy, the Queen’s University Belfast, the Space Telescope Science Institute, the South African Astronomical Observatory, and The Millennium Institute of Astrophysics (MAS), Chile.

The PS1 and the PS1 public science archives have been made possible through contributions by the Institute for Astronomy, the University of Hawaii, the Pan-STARRS Project Office, the Max-Planck Society and its participating institutes, the Max Planck Institute for Astronomy, Heidelberg and the Max Planck Institute for Extraterrestrial Physics, Garching, the Johns Hopkins University, Durham University, the University of Edinburgh, the Queen's University Belfast, the Harvard-Smithsonian Center for Astrophysics, the Las Cumbres Observatory Global Telescope Network Incorporated, the National Central University of Taiwan, the Space Telescope Science Institute, NASA under grant No. NNX08AR22G issued through the Planetary Science Division of the NASA Science Mission Directorate, NSF grant No. AST-1238877, the University of Maryland, Eotvos Lorand University, the Los Alamos National Laboratory, and the Gordon and Betty Moore Foundation.


\vspace{5mm}
\facilities{ADS, ATLAS, IRSA, FLWO (KeplerCam), LCO (MuSCAT3, Sinistro), MMT (Binospec), NED, ZTF}.

\defcitealias{AstropyCollaboration2018}{Astropy Collaboration 2018}
\software{Astropy \citepalias{AstropyCollaboration2018}, 
\texttt{atlas-fp} (\url{https://gist.github.com/thespacedoctor/86777fa5a9567b7939e8d84fd8cf6a76}), 
\texttt{barycorr} (\citealt{Eastman2010PASP..122..935E}; \url{https://github.com/tronsgaard/barycorr}),
\texttt{Cosimic-CONN} \citep{Xu2021arXiv210614922X,Xu2022arXiv220410356X,xu_chengyuan_2021_5034763},
\texttt{dustmaps} \citep{Green2018JOSS....3..695M}, 
\texttt{lcogtsnpipe} \citep{Valenti2016MNRAS.459.3939V}, 
Matplotlib \citep{Hunter2007}, 
NumPy \citep{Oliphant2006}, 
\texttt{PyZOGY} \citep{Guevel2017zndo...1043973G},
SciPy \citep{SciPy2020},
seaborn \citep{Waskom2020},
\texttt{SExtractor} \citep{Bertin1996}}. 


\bibliography{main}


\appendix
\restartappendixnumbering

\section{Optical Constraints for Well-localized Non-repeating FRBs}
\label{sec:norep}

\begin{figure*}
    \centering
    \includegraphics[width=0.925\textwidth]{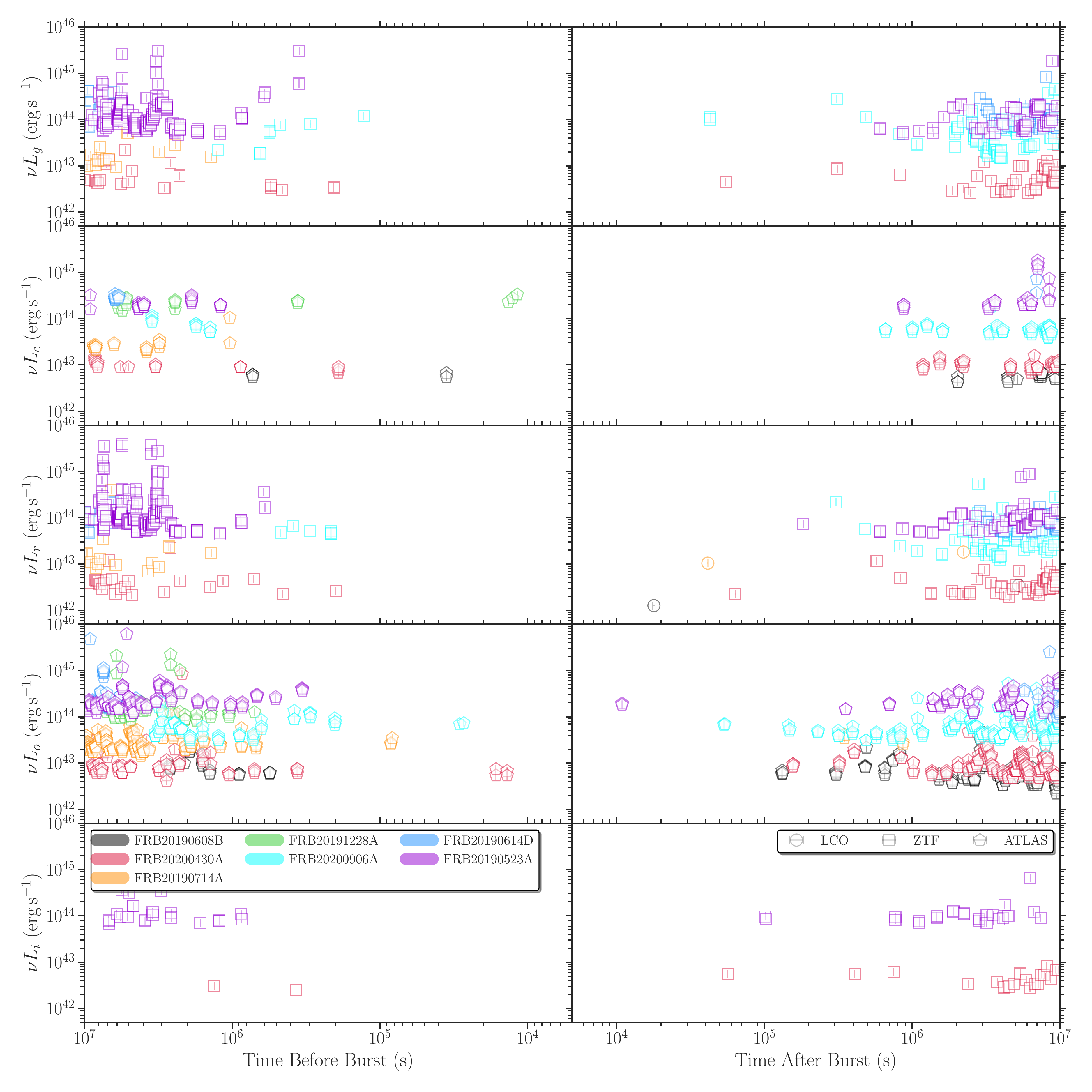}
\caption{Same as Figure~\ref{fig:lumlim}, but for the non-repeating FRBs (summarized in Table~\ref{tab:nonFRBsample}). Note the much less constraining luminosity and temporal ranges, compared to those of the repeating FRBs, given the small number of bursts and their large distances. Optical data sources: LCO (\citealt{Nunez2021AA...653A.119N} for FRBs~20190608B and 20190714A in the $r$ band), ZTF forced-photometry service (\citealt{Masci2019PASP..131a8003M} and this work for all the FRBs), ATLAS forced photometry server (\citealt{Shingles2021TNSAN...7....1S} and this work for all the FRBs). (The data used to create this figure are available.)} 
    \label{fig:nonlumlim}
\end{figure*}

\begin{figure*}
    \centering
    \includegraphics[width=0.925\textwidth]{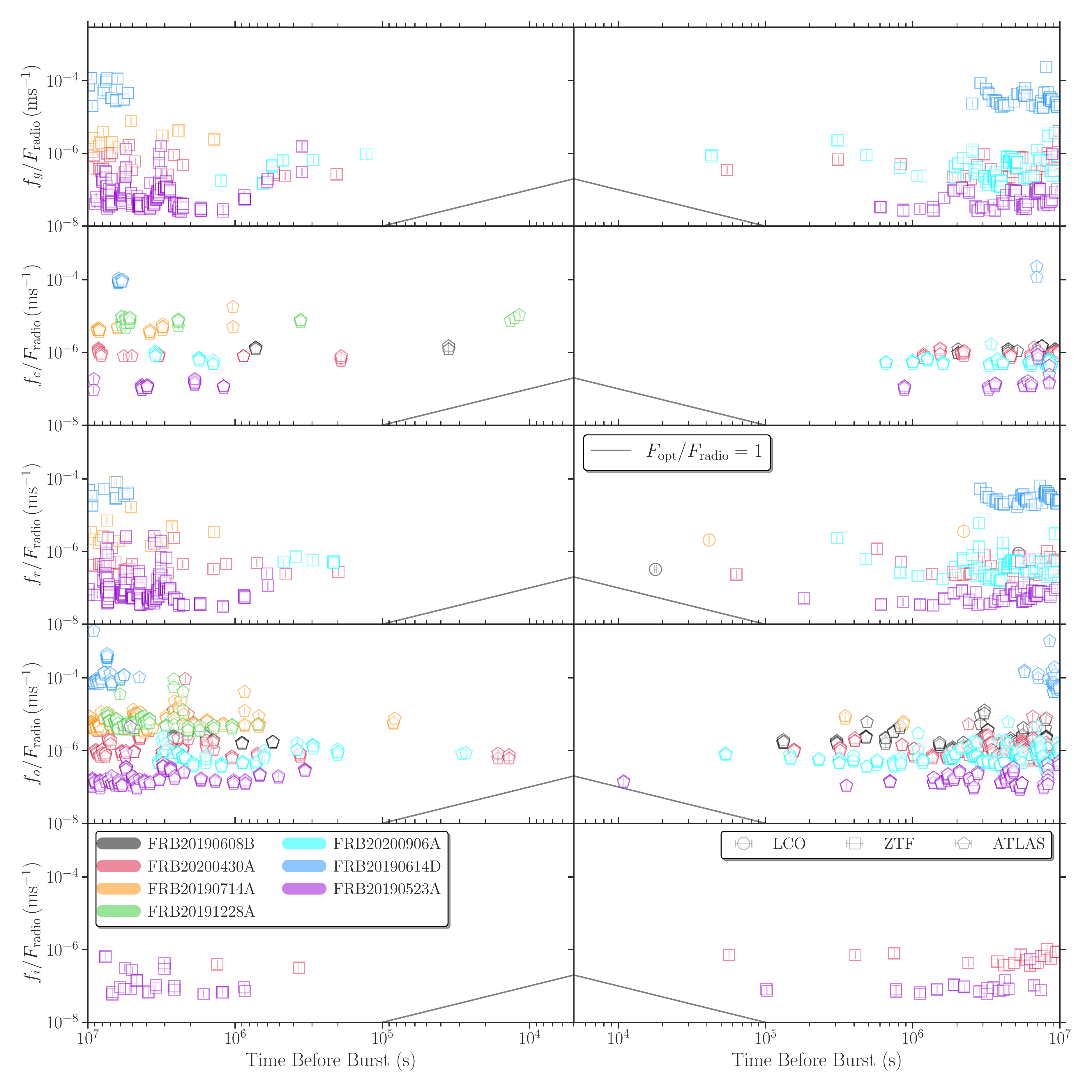}
\caption{Same as Figure~\ref{fig:ftoF}, but for the non-repeating FRBs. Note that all available observations are in $F_{\rm opt}/F_{\rm radio}\gtrsim 1$ regime.
    }
    \label{fig:nonftoF}
\end{figure*}

We summarize the well-localized non-repeating FRB sample and their references in Table~\ref{tab:nonFRBsample}. The same luminosity and flux analyses as for the repeating FRBs are performed and shown in Figures~\ref{fig:nonlumlim} and \ref{fig:nonftoF}, except for the simultaneous flux analysis since no such measurements are available given the unpredictable nature of one-off FRBs. We find that none of the optical limits are particularly constraining. Nevertheless, we also compare the luminosity limits to the light curves from the synchrotron maser model (although its application is somewhat questionable given the one-off nature of these events), concluding that only the very high-$E_{\rm flare}$ end ($\gtrsim10^{48}$\,erg) could be constrained. These constraints are at least three orders of magnitude larger than the FRB-based estimates (where available).

\begin{deluxetable*}{ccccccccccc}
\tabletypesize{\scriptsize}
\tablecaption{Well-localized Non-repeating FRB Sample\label{tab:nonFRBsample}}
\tablehead{
\colhead{FRB} & \colhead{R.A.} & \colhead{Decl.} & \colhead{Redshift} & \colhead{$dL$\tablenotemark{a}} & \colhead{$A_{V,{\rm MW}}$\tablenotemark{b}} & \colhead{Frequency} & \colhead{DM} & \colhead{Width\tablenotemark{c}} & \colhead{Flux\tablenotemark{c}} & \colhead{Fluence\tablenotemark{c}} \\[-6pt]
\colhead{} & \colhead{(deg)} & \colhead{(deg)} & \colhead{(z)} & \colhead{(Mpc)} & \colhead{(mag)} & \colhead{(GHz)} & \colhead{(pc\,cm$^{-3}$)} & \colhead{(ms)} & \colhead{(Jy)} & \colhead{(Jy\,ms)} 
}
\startdata
20190608B\tablenotemark{d} & 334.01988(1) & -7.89825(8) & $0.1178$ & $541$ & $0.125$  & $1.295$ & $338.7(5)$ & $6.0(8)$ &  $\sim4.3$ & $26(4)$ \\
20200430A\tablenotemark{e} & 229.70642(8) & +12.3769(3) & $0.160$ & $755$ & $0.112$ & $0.8645$ & $380.1(4)$ & -- & -- & $35(4)$ \\
20190714A\tablenotemark{f} & 183.9797(1) & -13.02103(8) & $0.2365$ & $1167$ & $0.166$ & $1.2725$ & $504(2)$ & -- & -- & $8(2)$ \\
20191228A\tablenotemark{g} & 344.4304(3) & -29.5941(3) & $0.2432$ & $1204$ & $0.065$ & $1.2715$ & $297.5(5)$ & $2.3(6)$ & $\sim17$ & $40^{+100}_{-40}$ \\
20200906A\tablenotemark{h} & 53.4962(1) & -14.0832(2) & $0.3688$ & $1946$ & $0.126$ & $0.8645$ & $577.8(2)$ & $6.0(6)$ & $\sim9.8$ & $59^{+25}_{-10}$ \\
20190614D\tablenotemark{i} & 65.07554(8) & +73.70675(8) & $0.6$ & $3480$ & $0.371$ & $1.4$ & $959(5)$ & $\sim5.0$ & $0.12(1)$ & $0.62(7)$ \\
20190523A\tablenotemark{j} & 207.0650(1) & +72.4697(6) & $0.660$ & $3908$ & $0.050$ & $1.530$ & $760.8(6)$ & $0.42(5)$ & $\gtrsim660$ & $\gtrsim280$ \\
\enddata

\tablecomments{Measurements of each individual burst are published in its entirety in machine-readable form.}

\tablenotetext{a}{Calculated from the host redshift assuming a standard $\Lambda$CDM cosmology with $H_0=71.0$\, km\,s$^{-1}$\,Mpc$^{-1}$, $\Omega_{\Lambda}=0.7$, and $\Omega_m=0.3$.}

\tablenotetext{b}{From \cite{Schlafly2011ApJ...737..103S}, retrieved via IRSA.}

\tablenotetext{c}{If only two of the width ($t_{\rm FRB}$), flux ($f_{\rm radio}$), and fluence ($F_{\rm radio}$) are reported for a particular burst, the third parameter is estimated assuming $F_{\rm radio} \sim f_{\rm radio}t_{\rm FRB}$.}

\tablenotetext{d}{Localized with ASKAP and the host spectroscopic redshift measured from an archival SDSS spectrum \citep{Day2020MNRAS.497.3335D,Macquart2020Natur.581..391M}.}

\tablenotetext{e}{Localized with ASKAP and the host spectroscopic redshift measured with Nordic Optical Telescope (NOT; \citealt{Heintz2020ApJ...903..152H, Kumar2020ATel13694....1K}).}

\tablenotetext{f}{Localized with ASKAP and the host spectroscopic redshift measured with Keck \citep{Bhandari2019ATel12940....1B, Heintz2020ApJ...903..152H}.}

\tablenotetext{g}{Localized with ASKAP and the host spectroscopic redshift measured with Keck \citep{Shannon2019ATel13376....1S, Bhandari2022AJ....163...69B}.}

\tablenotetext{h}{Localized with ASKAP and the host spectroscopic redshift measured with Keck \citep{Bhandari2022AJ....163...69B}.}

\tablenotetext{i}{Localized with VLA and the host photometric redshift estimated from spectral energy distribution fits \citep{Law2020ApJ...899..161L}.}

\tablenotetext{j}{Localized with DSA and the host spectroscopic redshift measured with Keck \citep{Ravi2019Natur.572..352R}.}

\end{deluxetable*}

\clearpage

\restartappendixnumbering

\section{Synchrotron Maser Model} 
\label{sec:mod}

With slight modifications, we follow \citet{Kilpatrick2021ApJ...907L...3K} for the application of the synchrotron maser model of \cite{Metzger2019MNRAS.485.4091M}, \cite{Margalit2020ApJ...899L..27M}, and \cite{Margalit2020MNRAS.494.4627M} to their optical limits on the well-localized repeating FRB~20180916B.
In this model, the prompt radio emission (i.e., FRB) and afterglow gamma-ray to optical emission originate from the shock interaction between magnetar/pulsar-ejected plasma and the highly magnetized external medium.
The characteristic synchrotron frequency ($\nu_{\rm syn}$) is set by the flare energy ($E_{\rm flare}$):
\begin{equation}
h \nu_{\rm syn}(t_{\rm FRB})=57\,{\rm MeV} \left( \frac{\sigma}{0.1} \right)^{1/2} \left( \frac{E_{\rm flare}}{10^{43}\,{\rm erg}} \right)^{1/2} \left( \frac{t_{\rm FRB}}{{\rm ms}} \right)^{-3/2},
\label{eq:nusync0}
\end{equation}
where $t_{\rm FRB}$ is the FRB duration (or width) and $\sigma$ is the fractional magnetization (see Equations (56) and (57) of \citealt{Metzger2019MNRAS.485.4091M}). In this work, we assume a fiducial $\sigma=0.3$ (\citealt {Plotnikov2019MNRAS.485.3816P}). The time evolution of the synchrotron frequency is as follows:
\begin{equation}
  \nu_{\rm syn}(t) =
    \begin{cases}
      \nu_{\rm syn}(t_{\rm FRB})\,\left(\frac{t}{t_{\rm FRB}}\right)^{-1} & \text{for $t < t_{\rm FRB}$}\\
      \nu_{\rm syn}(t_{\rm FRB})\,\left(\frac{t}{t_{\rm FRB}}\right)^{-3/2} & \text{for $t > t_{\rm FRB}$}.
    \end{cases}
\label{eq:nusync}
\end{equation}

\begin{figure*}
 \centering
 \gridline{\fig{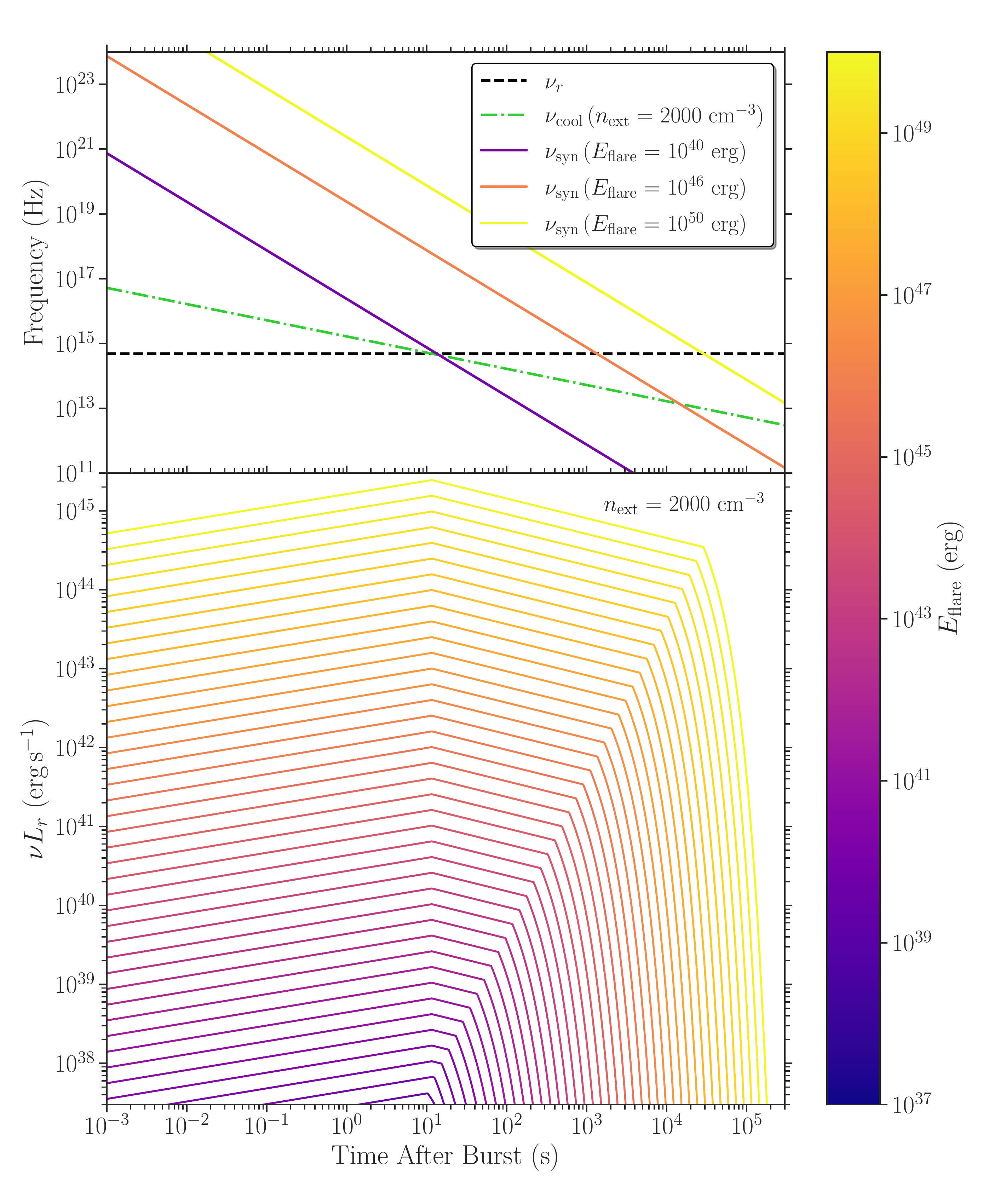}{0.475\textwidth}{}
          \fig{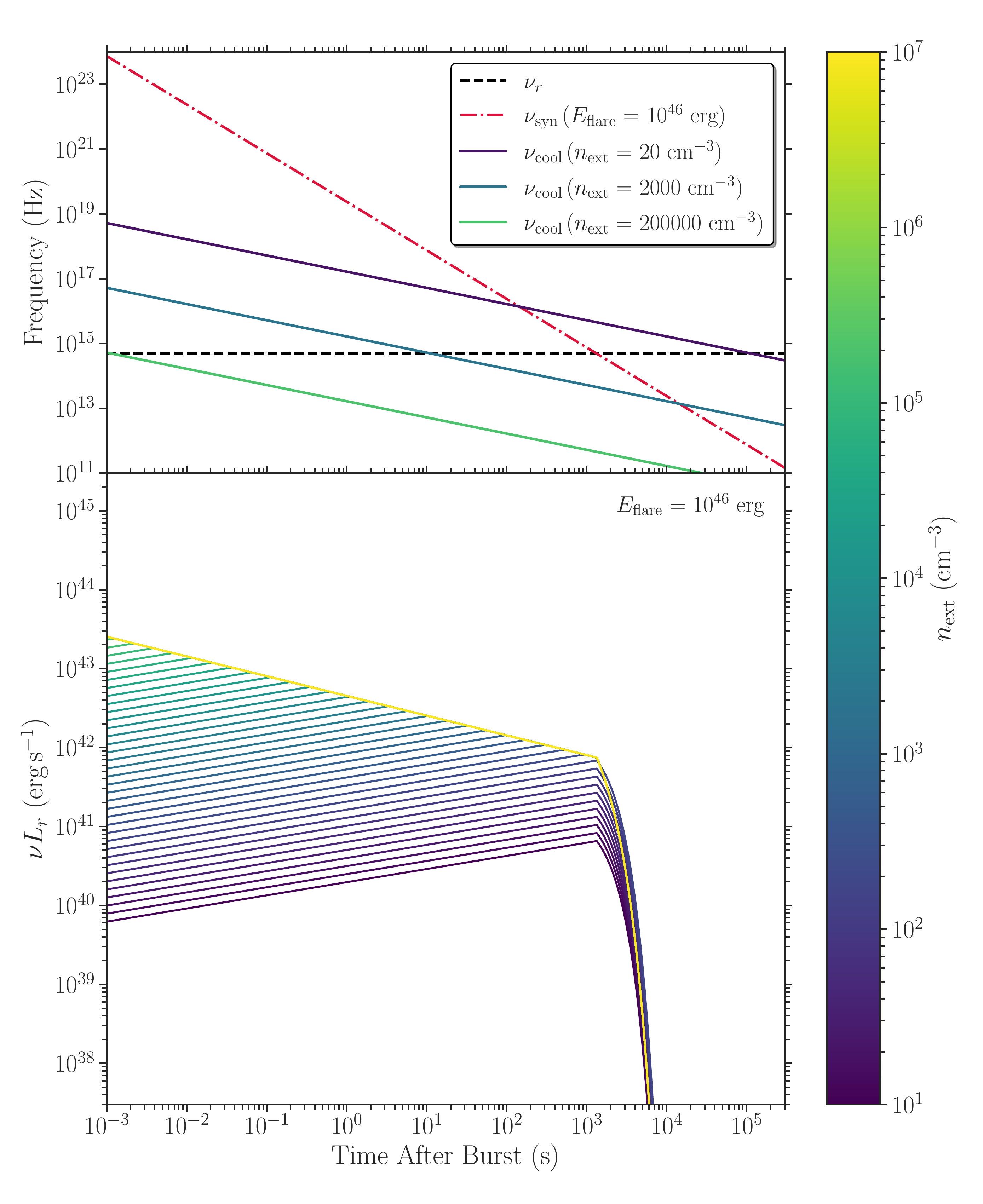}{0.475\textwidth}{}}
  \caption{Frequency and luminosity evolution (in the $r$ band) of the synchrotron maser model \citep{Metzger2019MNRAS.485.4091M,Margalit2020ApJ...899L..27M, Margalit2020MNRAS.494.4627M} with an FRB duration $t_{\rm FRB}=1$\,ms, assuming a shell-like density distribution and fractional magnetization $\sigma=0.3$ (see Equations (\ref{eq:nusync0})--(\ref{eq:nuLnu})). The light curves exhibit changing evolution in each frequency regime ($\nu_r<\nu_{\rm cool}$, $\nu_{\rm cool}<\nu_r<\nu_{\rm syn}$, and $\nu_{\rm syn}<\nu_r$). Left: models with a fixed density of $n_{\rm ext}=2000$\,cm$^{-3}$, color-coded by $E_{\rm flare}$. 
  Right: models with a fixed flare energy of $E_{\rm flare}=10^{46}$\,erg, color-coded by $n_{\rm ext}$.
  } 
  \label{fig:mod}
\end{figure*}

Assuming a constant (shell-like) external density ($n_{\rm ext}$) at the shock front, the cooling frequency ($\nu_{\rm cool}$) can be expressed as a function of time (see Equations (32) and (60) of \citealt{Metzger2019MNRAS.485.4091M} and also Equation (4) of \citealt{Kilpatrick2021ApJ...907L...3K}): 
\begin{equation}
  h\nu_{\rm cool}(t) = 2.3\,{\rm keV} \left(\frac{\sigma}{0.1}\right)^{-3/2} \left(\frac{n_{\rm ext}}{10^3\,{\rm cm}^{-3}}\right)^{-1} \left(\frac{t}{10^{-3}\,{\rm s}}\right)^{-1/2}.
\label{eq:nuc}
\end{equation}
Then, the peak luminosity ($L_{\rm peak}$) and specific luminosity ($L_\nu$) can be written as a function of frequency (see Equations (63) and (64) of \citealt{Metzger2019MNRAS.485.4091M}):
\begin{equation}
  L_{\rm peak}(t) = 10^{45}\,{\rm erg\,s}^{-1} \left(\frac{E_{\rm flare}}{10^{43}\,{\rm erg}} \right) \left(\frac{t}{10^{-3}\,{\rm s}}\right)^{-1},
\label{eq:Lpeak}
\end{equation}

\begin{equation}
  \nu L_\nu(t) =
    \begin{cases}
      L_{\rm peak} \left(\frac{\nu}{\nu_{\rm cool}}\right)^{4/3} \left(\frac{\nu_{\rm cool}}{\nu_{\rm syn}}\right)^{1/2} \propto t^{1/6} & \text{for $\nu < \nu_{\rm cool}$} \\
      L_{\rm peak} \left(\frac{\nu}{\nu_{\rm sync}}\right)^{1/2} \propto t^{-1/4} & \text{for $\nu_{\rm cool} < \nu < \nu_{\rm syn}$} \\
      L_{\rm peak} \, e^{-(\nu/\nu_{\rm syn}-1)} \propto t^{-1} \, e^{-t^{3/2}} & \text{for $\nu_{\rm syn} < \nu$} \\
    \end{cases}
\label{eq:nuLnu}
\end{equation}
where we adopt an exponential cutoff above $\nu_{\rm syn}$, as in \cite{Kilpatrick2021ApJ...907L...3K}. The time dependence is shown for $t>t_{\rm FRB}$. Representative model light curves in the $r$ band ($\nu=5\times10^{14}$\,Hz) for various choices of $E_{\rm flare}$ and $n_{\rm ext}$ are shown in Figure~\ref{fig:mod}.

From FRB measurements of the radio fluence ($F_{\rm radio}$) and frequency ($\nu_{\rm obs}$), the isotropic radio energy ($E_{\rm radio}$) and flare energy can be estimated as
\begin{equation}
  E_{\rm radio} = 9.5\times10^{31}\,{\rm erg} \left(\frac{4 \pi}{1+z}\right) \left(\frac{dL}{{\rm Mpc}}\right)^2 \left(\frac{F_{\rm radio}\cdot\nu_{\rm obs}}{{\rm Jy\,ms}\cdot{\rm GHz}}\right),
\label{eq:Eradio}
\end{equation}
\begin{equation}
  \frac{E_{\rm flare}}{E_{\rm radio}} = 3.3\times10^{4} \left(\frac{f_e}{0.5}\right)^{-1/5} \left(\frac{f_\xi}{10^{-3}}\right)^{-4/5} \left(\frac{\nu_{\rm obs} \cdot t_{\rm FRB}}{{\rm GHz \cdot ms}}\right)^{1/5},
\label{eq:Eflare}
\end{equation}
where $f_e$ and $f_\xi$ are the number density ratio of electrons to ions in the upstream medium
and the synchrotron maser efficiency, respectively (see Equation (14) of \citealt{Margalit2020ApJ...899L..27M}). In this work, we assume fiducial $f_e=0.5$ and $f_\xi=10^{-3}$ \citep{Plotnikov2019MNRAS.485.3816P}. The external density can also be estimated from FRB measurements as
\begin{equation}
\begin{split}
  n_{\rm ext} & = 420\,{\rm cm}^{-3}  \left(\frac{m_\star}{m_e}\right)^{2/15} \left(\frac{f_e}{0.5}\right)^{-11/15} \left(\frac{f_\xi}{10^{-3}}\right)^{-4/15} \\ 
  & \times \left(\frac{\nu_{\rm obs}}{{\rm GHz}}\right)^{31/30} \left( \frac{t_{\rm FRB}}{{\rm ms}} \right)^{2/5} \left(\frac{E_{\rm radio}}{10^{40}{\rm erg}}\right)^{-1/3},
\end{split}
\label{eq:next}
\end{equation}
where $m_\star$ and $m_e$ are the masses of upstream particles and electrons (see Equation (8) of \citealt{Margalit2020MNRAS.494.4627M}). In this work, we assume a pair plasma (i.e., $m_\star=m_e$).
Then, these FRB-based estimates for $E_{\rm flare}$ and $n_{\rm ext}$ can be compared with the optical constraints, as in Figure~\ref{fig:const}.

\end{document}